\documentclass[12pt]{article}

\begin{document}

{\noindent {\small CITUSC/01{-011} \hfill \hfill hep-th/0104135 \newline
}} \noindent {\small SNUST/010102 \hfill \newline
} \noindent {\small IHES/P/01/10 \hfill \newline
} \noindent {\small IHP/2001/18 \hfill \newline
}

{\vskip-0.8cm}

\begin{center}
{\Large \textbf{Noncommutative Sp(2,R) Gauge Theories \\[0pt]
A Field Theory Approach to Two-Time Physics}}

{\vskip0.4cm}

\textbf{Itzhak Bars}$^{a}$ \textrm{and} \textbf{Soo-Jong Rey}$^{b,c,d}$

{\vskip0.3cm}

$^{a)}$\textsl{CIT-USC Center for Theoretical Physics \& Department of
Physics}

\textsl{University of Southern California,\ Los Angeles, CA 90089-2535 USA}

{\vskip0.25cm}

$^{b)}$\textsl{School of Physics \& Center for Theoretical Physics}

\textsl{Seoul National University, Seoul 151-747 KOREA}

{\vskip0.25cm}

$^{c)}$ \textsl{Centre Emil Borel, Institut Henri Poincar\'e}

\textsl{11, rue Pierre et Marie Curie, Paris F-75231 FRANCE}

{\vskip0.25cm}

$^{d)}$ \textsl{Institut des Hautes \`Etudes Scientifiques}

\textsl{Le Bois-Marie, 35 route de Chartes, Bures-sur-Yvette F-91440 FRANCE}

{\vskip0.5cm}
\centerline{\tt bars@usc.edu \hskip1cm sjrey@gravity.snu.ac.kr}
{\vskip0.6cm} \textbf{abstract}
\end{center}

Phase-space and its relativistic extension is a natural space for realizing
Sp(2,R) symmetry through canonical transformations. On a D$\times $2
dimensional covariant phase-space, we formulate noncommutative field
theories, where Sp(2,R) plays a role as either a global or a gauge symmetry
group. In both cases these field theories have potential applications,
including certain aspects of string theories, M-theory, as well as quantum
field theories. If interpreted as living in lower dimensions, these theories
realize Poincar\'{e} symmetry linearly in a way consistent with causality
and unitarity. In case Sp(2,R) is a gauge symmetry, we show that the
spacetime signature is determined dynamically as (D-2,2). The resulting
noncommutative Sp(2,R) gauge theory is proposed as a field theoretical
formulation of two-time physics: classical field dynamics contains all known
results of `two-time physics', including the reduction of physical spacetime
from D to (D-2) dimensions, with the associated `holography' and `duality'
properties. In particular, we show that the solution space of classical
noncommutative field equations put all massless scalar, gauge,
gravitational, and higher-spin fields in (D-2) dimensions on equal-footing,
reminiscent of string excitations at zero and infinite tension limits.

{\vskip0.2cm} \newpage \baselineskip=18pt

\section{Introduction}

In this paper, we construct $D\times 2$-dimensional noncommutative field
theories (NCFT) with symmetry group Sp$(2,R)\times $ SO$^{\ast }(D)$ and
study their properties. We consider Sp$\left( 2,R\right) $ either as a
global or as a gauge symmetry. Denoting $D\times 2$ dimensional coordinates
as $X_{\mu }^{\mathrm{M}}$, the group Sp$\left( 2,R\right) \times $ SO$%
^{\ast }(D)$ acts on the $\mu =1,2$ index as a doublet $\mathbf{(2,1)}$ and
on the M$=1,2,\cdots ,D$ index as a vector $\mathbf{(1,D)}$. A variety of
choices are possible for the signature $\eta ^{MN}$ in $D$ dimensions or the
corresponding real form of the SO${}^{\ast }(D)\equiv $SO$\left(
D-n,n\right) $ group, where $n$ is the number of timelike dimensions.
Theories with Euclidean signature arise most prominently in the
Weyl-Wigner-Moyal formulation \cite{weyl}-\cite{moyal} of nonrelativistic
quantum mechanics, and in the noncommutative sector of M-theory (when a
magnetic background field is present, with a maximum possible symmetry Sp$%
\left( 2,R\right) \times $SO$\left( 5\right) $).
For Lorentzian signature, $n\geq 1$, one can view the coordinates $X_{\mu }^{%
\mathrm{M}}$'s as labelling a relativistic quantum phase-space, $X_{1}^{%
\mathrm{M}}=X^{\mathrm{M}}$ and $X_{2}^{\mathrm{M}}=P^{\mathrm{M}},$
extending the Weyl-Wigner-Moyal formulation of the nonrelativistic quantum
mechanics \cite{weyl}\cite{wigner}\cite{moyal} to relativistic situations.
One can also view them as noncommuting spacetime coordinates of two
point-particles, where the noncommutativity is induced by the presence of a
background field. Because vastly different interpretations are possible,
more broadly, we expect the formalism and methods developed for these
noncommutative field theories to have a wide range of applications in
various physical contexts.

Our main results for these noncommutative field theories are twofold:

\begin{itemize}
\item  If Sp$\left( 2,R\right) $ is a gauge symmetry, the spacetime
signature is determined dynamically as $(D-2,2)$. The gauge-invariant sector
of these theories describes commutative dynamics in $(D-2)$-dimensional
spacetime, with $\left( D-3,1\right) $ signature, with linearly realized
Poincar\'{e} and, non-linearly realized, higher spacetime symmetries. They
offer an \textsl{ab initio} field theory formulation of `Two-Time Physics'
(2T-physics) \cite{survey2T}-\cite{highspin} in $D$-dimensional spacetime, a
result established by interpreting $X_{\mu }^{\mathrm{M}}$ as coordinates of
the $D\times 2$ dimensional covariant phase-space.

\item  If Sp$\left( 2,R\right) $ is a global symmetry, the spacetime
signature is left arbitrary. Unitarity and causality force an interpretation
of these theories as describing dynamics in spacetime of dimensions lower
than $D$.
\end{itemize}

In investigating this sort of noncommutative field theories, the general
question we have posed ourselves is the following. Quantum field theories
are traditionally formulated on configuration space. Alternatively, what if
one attempts to formulate the theories on corresponding phase-space? In the
disguise of non-relativistic quantum mechanics, precisely this sort of an
alternative formulation was proposed by Weyl \cite{weyl}, Wigner \cite
{wigner}, and studied further by Moyal \cite{moyal}. It is referred as
`deformation quantization', an alternative to the traditional quantization
based on Hilbert space and linear operators therein. In this approach,
dynamical equations of quantum mechanics, either the Schr\"{o}dinger or the
Heisenberg equation of motion, are replaced by a sort of evolution equation
of distribution functions over phase-space. Mathematically, the
Weyl-Wigner-Moyal formalism is equivalent to noncommutative field theories
arising as a limit of string theories \cite{douglas1}-\cite{seibergwitten},
and is identifiable as the Euclidean case, $n=0$, in the present context.

Extension of the deformation quantization to systems with $n\geq 1$ poses
several peculiarities, all of which lead to a link with two timelike
dimensions. Lorentz covariance SO${}^{\ast }\left( D\right) $ implies that
time and energy (one-particle Hamiltonian) ought to be included, along with
spatial coordinates and momenta, as part of covariant phase-space. If so,
general canonical transformations consistent with Sp$\left( 2,R\right)
\times $SO${}^{\ast }\left( D\right) $ would require the one-particle
Hamiltonian to transform along with coordinates and momenta, thus mapping
one system with a given one-particle Hamiltonian to another with a different
one-particle Hamiltonian. This is a central feature that, for $n=2$,
2T-physics has embodied through local Sp$\left( 2,R\right) $, leading to the
`2T to 1T holography', and `duality' property among various systems with one
physical time \cite{bars}\cite{survey2T}. Apparently, a relativistic field
theory with more than one timelike dimensions introduces ghosts, as is
easily seen by expanding a field $\phi \left( X,P\right) $ in powers of
momenta whose coefficients are local tensors $\phi ^{\mathrm{M_{1}\cdots
M_{s}}}\left( X\right) $. Timelike components of these tensors could give
rise to negative-norm, ghost fields. An approach for eliminating the ghosts
is to promote Sp$(2,R)$ to a gauge symmetry, viz. by demanding equivalence
under general canonical transformations. The vanishing of the Sp$\left(
2,R\right) $ gauge generators on physical states determines dynamically that
spacetime must have two timelike dimensions.

It turns out that the resulting noncommutative Sp$(2,R)$ gauge theories
possess a rich structure, most notably the `holography' property, in which a
$D$-dimensional system is holographically represented by various $(D-2)$%
-dimensional systems, each with different dynamics. The signature of the $%
(D-2)$-dimensional spacetime is $\left( D-3,1\right) $, where the timelike
direction in each $\left( D-3,1\right) $-dimensional system is given by a
combination of the two timelike dimensions in the embedding $(D-2,2)$%
-dimensional spacetime. With one timelike dimension, the $(D-2)$-dimensional
systems are causal and have a unitary spectrum of physical states. The Sp$%
(2,R)$ gauge symmetry acts as a sort of `duality' in that all these
different dynamical systems are included in the Sp$(2,R)$ gauge orbit that
describes the physical gauge invariant sector.

This paper is organized as follows. In section 2, for later application, we
explain conceptual issues in constructing field theories of 2T-physics and
recapitulate some results of earlier approaches relevant for later sections.
In section 3, we discuss deformation quantization on covariant phase-space
and develop a formalism that will be used in later sections. In section 4,
we construct examples of noncommutative field theories with global Sp$%
(2,R)\times $ SO${}^{\ast }(D)$ symmetry for a generic signature of
spacetime. Results of these two sections are more general and ought to be
applicable to a wide range of physical problems. In section 5, we promote
the Sp$(2,R)$ automorphism to a gauge symmetry and construct noncommutative
Sp$(2,R)$ gauge theories. We show that, the condition for physical states
dynamically determines that the signature of spacetime must be $\left(
D-2,2\right) ,$ so inevitably we end up with 2T-physics. We then find a
class of nontrivial special classical solutions that reproduce all
previously known results of 2T-physics for spinless particles and, most
notably, find that the solution space provides a unified description of
gauge fields, including the gravitational and high spin gauge fields.
Section 6 summarizes various issues left for future work.

\section{2T-Physics: Concepts and Field Theory}

Part of the motivation for the present work has arisen from the following
question: What is the interacting field theory, whose free propagation is
given by the first quantized worldline theory of 2T-physics \cite{survey2T}-
\cite{highspin}? Free field equations emerge, in covariant first-quantized
description, by imposing constraints on states in configuration space, e.g.
worldline reparametrization constraints leads to the Klein-Gordon equation $%
\left[ \partial ^{2}-m^{2}\right] \phi (X)=0$, worldsheet reparametrization
constraints lead to string field equations through the Virasoro constraints $%
L_{n}\phi \left( X\left( \sigma \right) \right) $ = 0. These constraints are
the generators of the underlying gauge symmetries and hence the states
obeying them are gauge invariant physical states. In several known
situations, the constraint equations can be derived from a field theory when
the field interactions are neglected. Field interactions promote the first
quantized theory to an interacting field theory which can then be analyzed
both with classical and second quantized methods.

In 2T-physics, the fundamental gauge symmetry is Sp$\left( 2,R\right) $ and
its supersymmetric generalizations. Sp$\left( 2,R\right) $ acts as
symplectic transformation on coordinates and momenta of a particle's
phase-space $\left( X^{\mathrm{M}},P^{\mathrm{M}}\right) \equiv X_{\mu }^{%
\mathrm{M}}$. For a spinless particle, the worldline action with \textsl{%
local} Sp$\left( 2,R\right) $ symmetry is given by
\begin{equation}
I=\int d\tau \left[ \dot{X}_{1}^{\mathrm{M}}X_{2\mathrm{M}}-{\frac{1}{2}}%
A^{\mu \nu }\left( \tau \right) \,\hat{Q}_{\mu \nu }(X_{1},X_{2})\right] ,
\label{worldline}
\end{equation}
where, $A_{\mu \nu }\left( \tau \right) $ denotes three Sp$\left( 2,R\right)
$ gauge fields and the symmetric $\hat{Q}_{\mu \nu }=\hat{Q}_{\nu \mu },$
with $\mu ,\nu =1,2,$ are the three Sp$\left( 2,R\right) $ generators, whose
Poisson brackets obey sp$\left( 2,R\right) $ Lie algebra. This action is Sp$%
\left( 2,R\right) $ gauge invariant provided the $\hat{Q}_{\mu \nu }$
satisfy the Sp$\left( 2,R\right) $ Lie algebra under Poisson brackets \cite
{bars}\cite{highspin}. The equations of motion for $A^{\mu \nu }$ lead to
three classical constraints, $\hat{Q}_{\mu \nu }\left( X,P\right) =0$, which
become, upon first quantization, differential operator equations,
\begin{equation}
\hat{Q}_{\mu \nu }\left( X_{1},\frac{-i\partial }{\partial X_{1}}\right)
\psi (X_{1})=0,  \label{fieldeqs}
\end{equation}
with an appropriate operator ordering. The simplest situation occurs for the
following form of the generators which we refer to as the ``free'' case
(omitting the hat symbol)
\begin{equation}
Q_{11}=X\cdot X,\quad Q_{12}=X\cdot P,\quad Q_{22}=P\cdot P.  \label{free}
\end{equation}
where the dot products, $X\cdot P=X^{M}P^{N}\eta _{MN},$ are constructed
using a flat metric $\eta _{MN}$ of unknown signature (which later is
dynamically determined to be $\left( D-2,2\right) $). For the particle in
background fields such as Yang-Mills, gravity, higher-spin gauge fields, $%
\hat{Q}_{\mu \nu }\left( X,P\right) $ takes a more general form \cite{emgrav}
\cite{highspin}. With this example, one can see that the $\hat{Q}_{22}$%
-equation in (\ref{fieldeqs}) is nothing but (a generalization of) the
massless Klein-Gordon equation. In fact, the (22) components of the Sp$%
\left( 2,R\right) $ gauge fields and generators are associated with the
worldline reparametrization invariance. Together with the additional
transformations generated by $\hat{Q}_{11}$ and $\hat{Q}_{12},$ worldline
reparametrization is promoted to the non-Abelian local Sp$\left( 2,R\right) $
symmetry, which may also be regarded as local conformal symmetry on the
worldline \cite{bars}.

One of the goals of this work is to construct interacting field theories
with Sp$\left( 2,R\right) $ gauge symmetry, which would yield the first
quantized 2T-physics physical state equations (\ref{fieldeqs}) from the
linearized part of the field equations of motion. The obvious benefit of
these theories is, of course, to reach a formulation of field interactions
from first principles based on gauge symmetry. An interesting outcome of
this approach, as will be elaborated in Section 5, viz. noncommutative Sp$%
\left( 2,R\right) $ gauge theories, is that all massless fields in $(D-2)$
dimensions, scalar, gauge, gravity, and higher-spin gauge fields, are all
packaged into the noncommutative Sp$\left( 2,R\right) $ gauge field $A_{\mu
\nu }\left( X,P\right) $ in a $D\times 2$-dimensional covariant phase-space!
There seems to be an intriguing relationship between this packaging of
higher-spin gauge fields and a subset of massless string modes at infinite
Regge slope (tensionless string) or zero Regge slope (point-particle) limits
\cite{highspin}. The noncommutative Sp(2, R) gauge theories offers an
approach for a 2T-physics description of fields and the formulation of
nonlinear interactions among themselves.

The notion of two timelike dimensions raises various technical and
conceptual questions and points to deeper physics. Remarkably, the most
obvious and troublesome problem concerning causality and unitarity is
solvable, and transparently understood in the worldline approach to
2T-physics. The reason for two timelike dimensions is as follows. The Sp$%
\left( 2,R\right) $ gauge invariance imposes constraints, $Q_{\mu \nu
}\approx 0$, viz. physical states are defined as gauge singlets. Solving
them classically, one finds that nontrivial dynamics is possible only for
two or more timelike dimensions. However, the Sp$\left( 2,R\right) $ gauge
symmetry can remove all the ghosts only if the number of timelike dimensions
do not exceed two \footnote{%
As an illustration, take the simplest case Eq.(\ref{free}) wherein the inner
products $X\cdot X=X^{\mathrm{M}}X^{\mathrm{N}}\eta _{\mathrm{MN}},$ etc.
are defined with a flat metric of unknown signature. For Euclidean metric,
the only solution is a trivial one, $X^{\mathrm{M}}=P^{\mathrm{M}}=0$. For
Lorentzian metric with a single timelike dimension, $X^{\mathrm{M}}$ and $P^{%
\mathrm{M}}$ ought to be parallel, and is a trivial system since it lacks
angular momentum. For Lorentzian metric, with more than two timelike
dimensions, the Sp$\left( 2,R\right) $ gauge invariance is insufficient to
remove all the ghosts. For Lorentzian metric, with precisely two timelike
dimensions, Sp$\left( 2,R\right) $ gauge invariance is just enough to remove
all the ghosts. Furthermore, causality is not violated as, in a unitary
gauge, there is only one timelike dimension.}. Hence, Sp$\left( 2,R\right) $
gauge invariance demands two timelike dimensions, no less and no more.

A similar analysis is applicable in the present context and will lead
precisely to the same conclusion: noncommutative Sp$\left( 2,R\right) $
gauge theories on a $D\times 2$-dimensional phase-space with signature $%
\left( D-2,2\right) $ are unitary (as there are precisely two timelike
dimensions) and causal (as, in unitary gauge, the physical spacetime is
reduced to $(D-2)$ dimensions with one timelike dimension). In particular, $%
(D-2)$-dimensional spacetime symmetries -- Galilean, Poincar\'{e},
conformal, or even general coordinate invariance -- arise as symmetries.

If 2T-physics with Sp$(2,R)$ gauge symmetry is equivalent to one-time (1T)
physics, what does one gain from the former formulation? By embedding
1T-physics into 2T-physics, one is led to a notion of `holography' between
2T- and 1T-physics: a given 2T action in $D$ dimensions describes a family
of 1T actions in $(D-2)$ dimensions. Examples displaying these properties
have been found in worldline and field theory formulations of 2T-physics
\cite{survey2T}\cite{bars}\cite{field2T}. The `holography' is intimately
related to the issue of `time': which (combination) of the two times
corresponds to the causal evolution parameter of the physical 1T systems? In
the worldline formulation one can rephrase this question as: which
combination of $X^{0}\left( \tau \right) ,X^{0^{\prime }}\left( \tau \right)
$ is identified with the proper time $\tau $? It turns out that Sp$(2,R)$
gauge orbit in the physical sector, $\hat{Q}_{\mu \nu }\left( X,P\right) =0,$
encompasses all possible combinations. Furthermore, the Sp$(2,R)$ gauge
symmetry thins out the spacetime degrees of freedom from $D$ to $(D-2)$
dimensions, giving rise to the holography property. Thus, different 1T
theories in $(D-2)$ dimensions emerge as a result of different choices of
the Sp$(2,R)$ gauge fixing, but they all represent the physical sector of $D$
dimensional 2T-theory. This also implies that, within a given 2T-theory,
different 1T-theories are related to one another via a sort of `duality':
the Sp$(2,R)$ gauge transformations map a 1T-theory to another, while
staying within the physical (gauge-invariant) sector of the same 2T-theory.
The `holography' and `duality' properties ought to persist in noncommutative
Sp$(2,R)$ gauge theories, but now accommodating nonlinear gauge interactions.

We find it compelling to understand the above phenomena in a field-theoretic
formulation of 2T-physics, including interactions. An first attempt would be
in terms of fields defined on configuration space, as studied in \cite
{field2T}. However, it became clear that a more natural and far reaching
approach would result from a phase-space formulation. Naturally, the
resulting formulation is in terms of noncommutative Sp$(2,R)$ gauge
theories, which as shown below makes contact with the relevant parts of the
configuration space approach. Hence it is useful for us to review here the
salient aspects of the configuration space formulation \cite{field2T}.

Field equations in configuration space (in the presence of background
fields) result from imposing the constraints on physical states as in Eq.(%
\ref{fieldeqs}). For the free case of Eq.(\ref{free}) these take the form
\begin{equation}
\mathbf{q}_{\mu \nu }\psi \left( X_{1}\right) =0,  \label{first1}
\end{equation}
where $Q_{\mu \nu }\rightarrow \mathbf{q}_{\mu \nu }$ refers to hermitian
differential operators
\begin{equation}
\mathbf{q}_{11}=X_{1}\cdot X_{1},\quad \mathbf{q}_{12}=-\frac{i}{2}\left(
X_{1}\cdot \frac{\partial }{\partial X_{1}}+\frac{\partial }{\partial X_{1}}%
\cdot X_{1}\right) ,\quad \mathbf{q}_{22}=-\frac{\partial }{\partial X_{1}}%
\cdot \frac{\partial }{\partial X_{1}}.  \label{first2}
\end{equation}
The $\mathbf{q}_{11}$ equation is solved by
\begin{equation}
\psi \left( X_{1}^{\mathrm{M}}\right) =\delta \left( X_{1}^{2}\right)
\,\varphi \left( X_{1}^{\mathrm{M}}\right) .  \label{physical0}
\end{equation}
The $\mathbf{q}_{12}$ equation becomes
\begin{eqnarray}
0 &=&\mathbf{q}_{12}\,\psi \left( X_{1}^{\mathrm{M}}\right) =-i\left( \frac{D%
}{2}+X_{1}\cdot \frac{\partial }{\partial X_{1}}\right) \psi \left( X_{1}^{%
\mathrm{M}}\right)  \nonumber \\
&=&-i\delta \left( X_{1}^{2}\right) \left[ \left( \frac{D}{2}-2+X_{1}\cdot
\frac{\partial }{\partial X_{1}}\right) \,\varphi \left( X_{1}^{\mathrm{M}%
}\right) \right] _{X_{1}\cdot X_{1}=0},  \label{physical1}
\end{eqnarray}
where we have used an identity $X_{1}\cdot \frac{\partial }{\partial X_{1}}%
\delta \left( X_{1}^{2}\right) =2X_{1}^{2}\delta ^{\prime }\left(
X_{1}^{2}\right) =-2\delta \left( X_{1}^{2}\right) $ (as a distribution).
The $\mathbf{q}_{22}$ equation becomes
\begin{equation}
\left( \frac{1}{2}\mathbf{l}^{\mathrm{MN}}\mathbf{l}_{\mathrm{MN}}\right)
\,\left. \varphi \left( X_{1}^{\mathrm{M}}\right) \right| _{X_{1}\cdot
X_{1}=0}=-\frac{1}{4}D\left( D-4\right) \left. \varphi \left( X_{1}^{\mathrm{%
M}}\right) \right| _{X_{1}\cdot X_{1}=0}.  \label{physical2}
\end{equation}
Here, $\frac{1}{2}\mathbf{l}^{\mathrm{MN}}\mathbf{l}_{MN}$ is the is the SO$%
^{\ast }\left( D\right) $ quadratic Casimir operator and $\mathbf{l}^{MN}$
is its generator
\begin{equation}
\mathbf{l}^{\mathrm{MN}}=-i\left( X_{1}^{\mathrm{M}}\frac{\partial }{%
\partial X_{\mathrm{1N}}}-X_{1}^{\mathrm{N}}\frac{\partial }{\partial X_{%
\mathrm{1M}}}\right) .  \label{rot}
\end{equation}
Equation (\ref{physical2}) is a rewriting of $\frac{1}{2}\mathbf{q}_{\mu \nu
}\mathbf{q}^{\mu \nu }\varphi =\frac{1}{2}\left( \mathbf{q}_{11}\mathbf{q}%
_{22}+\mathbf{q}_{22}\mathbf{q}_{11}-2\mathbf{q}_{12}\mathbf{q}_{12}\right)
\varphi =0$ after using the relation between the Sp$\left( 2,R\right) $ and
the SO$\left( D-2,2\right) $ Casimirs
\begin{equation}
\frac{1}{2}\mathbf{l}^{\mathrm{MN}}\mathbf{l}_{\mathrm{MN}}=\frac{1}{2}%
\mathbf{q}_{\mu \nu }\mathbf{q}^{\mu \nu }-\frac{1}{4}D\left( D-4\right) ,
\label{spso}
\end{equation}
which is derived directly from their representations in Eqs.(\ref{first2},%
\ref{rot}). Thus demanding an Sp$\left( 2,R\right) $ gauge invariant
physical state Eq.(\ref{first1}) implies that such states form an
irreducible representation of SO$^{\ast }\left( D\right) $ with a fixed
eigenvalue for the quadratic Casimir operator of SO$^{\ast }\left( D\right) $
as given in Eq.(\ref{physical2}). The higher Casimir operators for SO$^{\ast
}\left( D\right) $ can be computed in the same way, and shown that they are
fixed numbers. Hence the physical states in the free case occupy a specific
representation of SO$^{\ast }\left( D\right) =$SO$\left(D-2,2\right) $. This
representation is a unitary representation, and are referred as \textsl{%
singleton} or \textsl{doubleton} representation, depending on the dimension $%
D$.

The above differential equations Eqs.(\ref{physical0}-\ref{physical2}) have
non-trivial solutions only if there are two timelike dimensions. Moreover,
the particular SO$^{\ast }\left( D\right) $ representation emerging in this
way is unitary provided there are again two timelike dimensions. This
implies that the first-quantized theory requires SO$^{\ast }\left( D\right)
= $SO$\left( D-2,2\right) $ with two timelike dimensions, confirming the
result of the classical analysis recapitulated earlier.

The holographic aspects can be studied in the 2T-field theory. One
holographic picture is the $(D-2)$-dimensional massless Klein-Gordon
equation, derived originally by Dirac \cite{Dirac}. Another is the
nonrelativistic Schr\"{o}dinger equation, and yet another is the scalar
field equation in anti-de Sitter background with a quantized mass, etc. \cite
{field2T}. In each of them, the SO$\left( D-2,2\right) $ automorphism of the
2T-theory arises with different physical interpretation. It is interpreted
as conformal symmetry of the Klein-Gordon equation, while, for others, as
dynamical symmetry or anti-de Sitter symmetry etc. The existence of this
symmetry in some of the 1T-theories is surprising, but it is understood
naturally within the 2T-framework. Furthermore, all 1T holographic pictures
of the free 2T-physics theory (free massless particle, AdS$_{d}$ particle,
AdS$_{d-k}\times $S$^{k}$ particle, H-atom, Harmonic oscillator in one less
dimension, etc.) occupy the same singleton/doubleton representation
described above \cite{bars}.

Generalizations of the same approach to field theory, including field
interactions, and including spinning particles, gauge, and gravitational
fields, etc. were accomplished \cite{field2T}.

However, one unsatisfactory aspect is that the equations $\mathbf{q}_{\mu
\nu }\psi =\cdots $ are not all treated on an equal footing: the $\mathbf{q}%
_{22}$ condition, including interactions, is derivable from an interacting
2T-theory action, however, the $\mathbf{q}_{11}$ and $\mathbf{q}_{12}$
conditions do not follow directly from the action and are applied as
additional constraints (although one could introduce them by using Lagrange
multipliers). One thus anticipates \cite{bars} that 2T-field theories ought
to be constructed most naturally as noncommutative field theories on the
phase-space spanned by $\left( X_{1}^{\mathrm{M}},X_{2}^{\mathrm{M}}\right) $%
, as this is the space where the Sp$\left( 2,R\right) $ transformations are
manifest, and all $Q_{\mu \nu }$ appear on an equal footing.

To construct noncommutative field theories that reproduce known results of
2T-physics, we will develop some formalism in the next two sections. We will
focus on how to maintain the Sp$\left( 2,R\right) \times $ SO$^{\ast }\left(
D\right) $ covariance manifest and study the theories in cases where Sp$%
\left( 2,R\right) $ symmetry is global or local. The Sp$\left( 2,R\right) $
gauge symmetry is the necessary ingredient for 2T-physics and leads to the
same results as the classical and the first-quantized 2T-theory. In
noncommutative field theories, however, the Sp$(2,R)$ gauge symmetry renders
consistent interactions as well. In the free field limit, field equations in
configuration space Eqs.(\ref{first1},\ref{first2}) follow naturally from
the noncommutative field equations. Solutions to these equations and their
holographic 1T interpretation coincide with the previous results \cite
{field2T} of the 2T field theory in configuration space. The noncommutative
field theories also yield known results when generic background fields are
turned on \cite{emgrav}\cite{highspin}. We thus find that the noncommutative
Sp$(2,R)$ gauge theories offer a unified approach to all aspects of
2T-physics, including interactions.

So far, we have considered mainly the phase-space interpretation of the
noncommuting coordinates $X_{1}^{\mathrm{M}}=X^{\mathrm{M}},$ $X_{2}^{%
\mathrm{M}}=P^{\mathrm{M}}$. On the other hand, as mentioned in the previous
section, we may also consider a noncommutative geometry interpretation of $%
X_{1}^{\mathrm{M}},X_{2}^{\mathrm{M}}$ as noncommuting \textit{positions }of
\textit{two} point-particles, where noncommutativity is induced by a
constant magnetic field. This idea naturally occurred in the context of a
two-particle system described by $\left( X_{1}^{\mathrm{M}},P_{1}^{\mathrm{M}%
}\right) $ and $\left( X_{2}^{\mathrm{M}},P_{2}^{\mathrm{M}}\right) ,$ with
constraints $P_{1}^{2}=P_{2}^{2}=P_{1}\cdot P_{2}=0$ that followed from
two-particle gauge symmetries \cite{multi}. In the presence of a constant
magnetic field with interactions $B\left( \dot{X}_{1}\cdot X_{2}-\dot{X}%
_{2}\cdot X_{1}\right) $ the two position coordinates $X_{1}^{\mathrm{M}%
},X_{2}^{\mathrm{M}}$ develop mutual non-commutativity. (In the infinite
magnetic field limit, the kinetic terms are negligible and then one may
reinterpret the system as a single particle, where $X_{1}=X,$ $X_{2}=P$ are
phase-space variables\footnote{%
The derivation of one particle dynamics from a two-particle system with
two-times, where non-commutativity is induced by a constant magnetic field,
was the historical path that led to the concepts in the first paper in \cite
{bars}.}). Such a set-up is analogous in spirit to the interpretation of
noncommutative field theories in terms of `dipole' behavior of an open
string theory in background magnetic field.


\section{$\star$--Algebra on Relativistic Quantum Phase-Space}

In this section we develop an \textit{ab initio} approach. Consider the
non-commutative (NC) Moyal products for any two functions $f\left( x\right)
\star g\left( x\right) $ in 2$\times $D dimensions. Instead of the generic
NC spacetime $x^{m}$ which satisfies $x^{m}\star x^{n}-x^{n}\star
x^{m}=i\theta ^{mn},$ we are interested in a special form of the
non-commutativity parameter $\theta ^{mn}$ that explicitly exhibits the
highest possible symmetry. Recalling that in a real basis for $x^{m}$ the
parameter $\theta ^{mn}$ may be brought to block diagonal form with skew $%
2\times 2$ blocks, the highest symmetry is manifest when all the $2\times 2$
diagonal blocks are identical up to signs. Such a $\theta ^{mn}$ parameter
has the symmetry Sp$^{\ast }\left( 2D\right) $ which contains the subgroup Sp%
$\left( 2,R\right) \times $SO$\left( D-n,n\right) $ for some $n.$ For
notational convenience we will write SO$^{\ast }\left( D\right) \equiv $SO$%
\left( D-n,n\right) $ and Sp$\left( 2\right) \equiv $Sp$\left( 2,R\right) .$
To exhibit this subgroup symmetry, it is convenient to use the pair of
labels $m=\mu \mathrm{M}$ with $\mu =1,2$ and M=1,2,$\cdots ,D,$ so that
spacetime is labelled by $X_{\mu }^{\mathrm{M}}$ instead of $x^{m},$ and $%
\theta ^{mn}$ is replaced by $\hbar \varepsilon _{\mu \nu }\eta ^{\mathrm{MN}%
},$ where $\varepsilon _{\mu \nu }$ and $\eta ^{\mathrm{MN}}$ are the
invariant metrics for Sp$\left( 2\right) $ and SO$^{\ast }\left( D\right) $
respectively. In this basis the Moyal $\star $-product takes the form
\begin{equation}
\left( f\star g\right) \left( X_{\mu }^{\mathrm{M}}\right) =\left. \exp
\left( \frac{i\hbar }{2}\varepsilon _{\lambda \sigma }\eta ^{\mathrm{MN}}%
\frac{\partial }{\partial X_{\lambda }^{\mathrm{M}}}\frac{\partial }{%
\partial \widetilde{X}_{\sigma }^{\mathrm{N}}}\right) f\left( X_{\mu }^{%
\mathrm{M}}\right) g\left( \widetilde{X}_{\mu }^{\mathrm{M}}\right) \right|
_{X_{\mu }^{\mathrm{M}}=\widetilde{X}_{\mu }^{\mathrm{M}}}.  \label{star}
\end{equation}
We define the $\star $-commutator
\begin{equation}
\Big[f(X),g(X)\Big]_{\star }:=f(X)\star g(X)-g(X)\star f(X).
\end{equation}
We then have the Heisenberg algebra:
\begin{equation}
\left[ X_{\mu }^{\mathrm{M}},X_{\nu }^{\mathrm{N}}\right] _{\star
}=i\varepsilon _{\mu \nu }\eta ^{\mathrm{MN}},
\end{equation}
which exhibits a global automorphism symmetry Sp$\left( 2, R\right) \times $%
SO$^{\ast }\left( D\right) .$ Hereafter we will set $\hbar =1$.

In the NC limit of 11-dimensional M-theory the highest such symmetry would
be Sp$\left( 2, R\right) \times $ SO$\left( 5\right) $ with Euclidean
signature. Our formalism would be useful in this physical setting. More
generally, concerning the spacetime signature, for now, we will take the
signature arbitrary, say, $(D-n)$ spacelike and $n$ timelike dimensions.
Ultimately, we shall be promoting the Sp$\left( 2, R\right) $ subgroup to a
gauge symmetry, and find that, as a consequence of the gauge invariance, the
number of timelike dimensions $n$ is determined uniquely to be $n=2$.

In this basis, there is no loss of generality if we consider the Sp$\left(
2, R\right) $ doublet $X_{\mu }^{\mathrm{M}}$ as the doublet of $D$%
-dimensional spacetime positions and energy-momenta: $X_{\mu }^{\mathrm{M}%
}=\left( X^{\mathrm{M}},P^{\mathrm{M}}\right) $, spanning $D\times 2$
dimensional relativistic phase-space. The subgroup SO$^{\ast }\left(
D\right) $ remains as a global subgroup of the relativistic phase-space.


\subsection{Symmetry Generators on Quantum Phase-Space}

Having identified the Sp(2)$\times $ SO${}^{\ast }(D)$ as the global
symmetry groups on the relativistic quantum phase-space, we now investigate
their Lie algebra, but in terms of the $\star $-product through the
Weyl-Moyal map. Denote the Sp$\left( 2\right) $ generators as $\mathbf{Q}%
_{\mu \nu }$ and the SO$^{\ast }\left( D\right) $ generators as $\mathbf{L}^{%
\mathrm{MN}}$, respectively. In terms of the $\star $-product, we have found
that they are represented by:
\begin{eqnarray}
\,\,\mathbf{Q}_{\mu \nu }\,\, &\equiv &\frac{1}{2}\eta _{\mathrm{MN}}X_{(\mu
}^{\mathrm{M}}\star X_{\nu )}^{\mathrm{N}}=\frac{1}{2}\eta _{\mathrm{MN}%
}\left( X_{\mu }^{\mathrm{M}}\star X_{\nu }^{\mathrm{N}}+X_{\nu }^{\mathrm{M}%
}\star X_{\mu }^{\mathrm{N}}\right) =\eta _{\mathrm{MN}}X_{\mu }^{\mathrm{M}%
}X_{\nu }^{\mathrm{N}},  \label{QandL} \\
\mathbf{L}^{\mathrm{MN}} &\equiv &\frac{1}{2}\varepsilon ^{\mu \nu }X_{\mu
}^{[\mathrm{M}}\star X_{\nu }^{\mathrm{N]}}=\frac{1}{2}\,\varepsilon ^{\mu
\nu }\,\left( X_{\mu }^{\mathrm{M}}\star X_{\nu }^{\mathrm{N}}-X_{\mu }^{%
\mathrm{N}}\star X_{\nu }^{\mathrm{M}}\right) =\,\varepsilon ^{\mu \nu
}\,X_{\mu }^{\mathrm{M}}X_{\nu }^{\mathrm{N}},  \label{Q&L}
\end{eqnarray}
where the symbols enclosed in parentheses or brackets, ($\mu \nu $), [MN]
etc., refer to symmetrization and antisymmetrization, respectively. The last
form of $\,\mathbf{Q}_{\mu \nu }\,,\,\ $after the star products have been
evaluated, is identical to the classical form of Eq.(\ref{free}). The same
remark applies to $\mathbf{L}^{\mathrm{MN}}.$ These, $\mathbf{Q}_{\mu \nu }$%
's and $\mathbf{L}^{\mathrm{MN}}$'s obey the sp(2)$\oplus $so${}^{\ast }(D)$
Lie algebras under star products
\begin{eqnarray}
\Big[\,\mathbf{Q}_{\mu \nu }\,,\,\mathbf{Q}_{\kappa \lambda }\,\Big]_{\star
} &=&i\,\,{F_{\mu \nu ,\kappa \lambda }}^{\alpha \beta }\,\,\,\mathbf{Q}%
_{\alpha \beta }  \label{spso2} \\
\left[ \mathbf{L}^{\mathrm{MN}}\,,\,\mathbf{L}^{\mathrm{KL}}\right] _{\star
} &=&i\,{F^{\mathrm{MN,KL}}}_{\mathrm{RS}}\mathbf{L}^{\mathrm{RS}}
\label{spso3} \\
\left[ \mathbf{L}^{\mathrm{MN}},\,\mathbf{Q}_{\mu \nu }\,\right] _{\star }
&=&\,0.  \label{spso1}
\end{eqnarray}
Here, ${F_{\mu \nu ,\kappa \lambda }}^{\alpha \beta },$ ${F^{\mathrm{MN,KL}}}%
_{\mathrm{RS}}$ denote the structure constants of the sp$\left( 2\right)
\oplus $ so$^{\ast }\left( D\right) $ Lie algebras, respectively:
\begin{equation}
{F_{\mu \nu ,\lambda \sigma }}^{\alpha \beta }=\frac{1}{2}\delta _{(\mu
}^{(\alpha }\,\,\varepsilon _{\nu )(\kappa }\,\,\delta _{\lambda )}^{\beta
)}\qquad \mathrm{and}\qquad {F^{\mathrm{MN,KL}}}_{\mathrm{RS}}=\frac{1}{2}%
\delta _{\lbrack R}^{[\mathrm{M}}\,\,\eta ^{\mathrm{N][K}}\,\,\delta _{%
\mathrm{S]}}^{\mathrm{L]}}. \label{structures}
\end{equation}

From the $\star $-product representation of the generators, we construct the
quadratic Casimir operators of Sp$\left( 2\right) $ and SO$^{\ast }\left(
D\right) $:
\begin{eqnarray}
C_{2}[\mathrm{SO^{\ast }\left( D\right) }] &=&\frac{1}{2}\mathbf{L}^{\mathrm{%
MN}}\star \mathbf{L}_{\mathrm{MN}},  \label{c2soD} \\
C_{2}[\mathrm{Sp\left( 2\right) }]\,\, &=&\frac{1}{2}\mathbf{Q}_{\mu \nu
}\star \mathbf{Q}^{\mu \nu },  \label{csp2}
\end{eqnarray}
where indices are contracted by using the metrics $\eta ^{MN}$ and $%
\varepsilon _{\mu \nu }$ respectively. Remarkably, by applying the $\star $%
-commutator relation Eq.(\ref{star}), one can show that the two Casimir
invariants are related each other just as in (\ref{spso})
\begin{equation}
C_{2}[\mathrm{SO^{\ast }\left( D\right) }]\,=\,C_{2}[\mathrm{Sp\left(
2\right) }]-\frac{1}{4}D\left( D-4\right) .  \label{L2Q2}
\end{equation}
Note that the relation is \textsl{independent} of the signature of the
D-dimensional spacetime. In the following discussions, Eq.(\ref{L2Q2}) will
play an important role, especially, in relating the resulting noncommutative
field theory to two-time physics.

\subsection{Differential Calculus on Quantum Phase-Space}

On the relativistic quantum phase-space, $\mathcal{M}_{\hbar }$,
differential calculus may be developed from the defining algebra of the $%
\star $-products. We thus consider left- or right-multiplication of single
power of $X_{\mu }^{\mathrm{M}}$'s from the left or the right of a function $%
\phi \left( X\right) $ on phase-space. They are:
\begin{eqnarray}
X_{\mu }^{\mathrm{M}}\star \phi (X) &=&\left( X_{\mu }^{\mathrm{M}}+\frac{i}{%
2}\partial _{\mu }^{\mathrm{M}}\right) \phi \left( X\right) \equiv \mathcal{D%
}_{\mu }^{\mathrm{M}}\phi (X)  \label{leftX} \\
\phi (X)\star X_{\mu }^{\mathrm{M}} &=&\left( X_{\mu }^{\mathrm{M}}-\frac{i}{%
2}\partial _{\mu }^{\mathrm{M}}\right) \phi \left( X\right) \equiv \overline{%
\mathcal{D}}_{\mu }^{\mathrm{M}}\phi (X).  \label{rightX}
\end{eqnarray}
Here, utilizing the invariant metrics $\varepsilon _{\mu \nu }$ and $\eta _{%
\mathrm{MN}}$, we have introduced the notation:
\begin{equation}
\partial _{\mu }^{\mathrm{M}}\equiv \eta ^{\mathrm{MN}}\varepsilon _{\mu \nu
}\frac{\partial }{\partial X_{\nu }^{\mathrm{N}}}\qquad \mathrm{such}\,\,\,%
\mathrm{that}\qquad \partial _{\mu }^{\mathrm{M}}X_{\nu }^{\mathrm{N}%
}=\varepsilon _{\mu \nu }\eta ^{MN}.  \label{der}
\end{equation}
The multiplications define, as the notations indicate, two inequivalent
differential operators -- $\mathcal{D}_{\mu }^{\mathrm{M}}\phi (X)$ and $%
\overline{\mathcal{D}}_{\mu }^{\mathrm{M}}\phi (X)$. However, these
differential operators violate the Leibniz rule: $\mathcal{D}_{\mu }^{%
\mathrm{M}}\left( \phi _{1}\star \phi _{2}\right)$ $\neq$ $\left( \mathcal{D}%
_{\mu }^{\mathrm{M}}\phi _{1}\right) \star \phi _{2}+\phi _{1}\star \left(
\mathcal{D}_{\mu }^{\mathrm{M}}\phi _{2}\right) $. On the other hand, a new
differential operator obeying the Leibniz rule can be defined by taking
\textsl{difference} between the above two differential operators:
\begin{equation}
\left( \mathcal{D}_{\mu }^{\mathrm{M}}-\overline{\mathcal{D}}_{\mu }^{%
\mathrm{M}}\right) \phi (X)=\left[ X_{\mu }^{\mathrm{M}},\phi (X)\right]
_{\star }\,=\,i\partial _{\mu }^{\mathrm{M}}\phi (X).  \label{deriv}
\end{equation}
The three differential operators, $\mathcal{D}_{\mu }^{\mathrm{M}},\overline{%
\mathcal{D}}_{\mu }^{\mathrm{M}},\partial _{\mu }^{\mathrm{M}}$, form a
complete set of first-order differential operators on the quantum
phase-space \footnote{%
More generally, one can construct a family of first-order differential
operators:
\[
\mathcal{D}_{1}^{\mathrm{M}}=\alpha X_{1}^{\mathrm{M}}+\beta \,\eta ^{%
\mathrm{MN}}\frac{i\partial }{\partial X_{2}^{\mathrm{N}}}\qquad \mathrm{and}%
\qquad \mathcal{D}_{2}^{\mathrm{M}}=\gamma X_{2}^{\mathrm{M}}-{\frac{1}{%
\alpha }}\left( 1-\beta \gamma \right) \eta ^{\mathrm{MN}}\frac{i\partial }{%
\partial X_{1}^{\mathrm{N}}}
\]
with arbitrary coefficients $\alpha ,\beta ,\gamma $. For $\alpha =\gamma =1$
and $\beta =1/2$, they reduce to Eqs.(\ref{leftX},\ref{rightX}). For $\alpha
=1$ and $\beta =\gamma =0$, they reduce to the conventional position and
momentum operators $\mathcal{D}_{1}^{\mathrm{M}}\equiv X^{\mathrm{M}}$ and $%
\mathcal{D}_{2}^{\mathrm{M}}\equiv P^{\mathrm{M}}=-i\partial /\partial X^{%
\mathrm{M}}$. We shall restrict the following discussion to the derivations
Eqs.(\ref{leftX},\ref{rightX},\ref{deriv}) only.}.

Next, consider $\star $-multiplication by two powers of $X_{\mu }^{\mathrm{M}%
}$'s on $\phi (X)$. Of particular interest are the generators, $\mathbf{Q}%
_{\mu \nu }$ and $\mathbf{L}^{\mathrm{MN}}.$ Their commutators define
derivatives that obey the Leibniz rule
\begin{eqnarray}
D_{\mu \nu }\phi (X) &\equiv &-i\,\left[ \mathbf{Q}_{\mu \nu },\phi \right]
_{\star }\,=\frac{1}{2}\eta _{\mathrm{MN}}\left( X_{(\mu }^{\mathrm{M}}\star
\partial _{\nu )}^{\mathrm{N}}\phi +\partial _{(\mu }^{\mathrm{M}}\phi \star
X_{\nu )}^{\mathrm{N}}\right) (X)=\eta _{\mathrm{MN}}X_{(\mu }^{\mathrm{M}%
}\partial _{\nu )}^{\mathrm{N}}\,\phi (X)  \label{Dij} \\
D^{\mathrm{MN}}\phi (X) &\equiv &-i\left[ \mathbf{L}^{\mathrm{MN}},\phi %
\right] _{\star }=\frac{1}{2}\varepsilon ^{\mu \nu }\left( X_{\mu }^{\mathrm{%
[M}}\star \partial _{\nu }^{\mathrm{N]}}\phi +\partial _{\mu }^{\mathrm{[M}%
}\phi \star X_{\nu }^{\mathrm{N]}}\right) (X)=\varepsilon ^{\mu \nu }X_{\mu
}^{\mathrm{[M}}\partial _{\nu }^{\mathrm{N]}}\,\phi (X).  \label{DMN}
\end{eqnarray}
Note the $\star $-multiplication ordering in the middle expressions. After
applying Eqs.(\ref{leftX},\ref{rightX}), however, they are expressible in
terms of ordinary products, as shown in the last expressions. Note further
that, using Eq.(\ref{der}), these two derivations can be expressed as total
first order derivatives:
\begin{equation}
D_{\mu \nu }\phi (X)=\partial _{(\mu }^{\mathrm{M}}\left( \eta _{\mathrm{MN}%
}X_{\nu )}^{\mathrm{N}}\phi \right) (X)\qquad \mathrm{and}\qquad D^{\mathrm{%
MN}}\phi (X)=\partial _{\mu }^{\mathrm{[M}}\left( \varepsilon ^{\mu \nu
}X_{\nu }^{\mathrm{N]}}\phi \right) (X),
\end{equation}
implying that integrals over the phase-space of these derivations acting on
smooth functions vanish identically.

Left-multiplications of the generators $\mathbf{Q}_{\mu \nu }$ and $\mathbf{L%
}^{\mathrm{MN}}$ on a function $\phi (X)$ define second-order differential
operators $\mathcal{D}_{\mu \nu },\mathcal{D}^{\mathrm{MN}}$:
\begin{eqnarray}
\mathbf{Q}_{\mu \nu }\,\star \phi (X)\, &=&\frac{1}{2}\eta _{\mathrm{MN}%
}\left( X_{(\mu }^{\mathrm{M}}+\frac{i}{2}\partial _{(\mu }^{\mathrm{M}%
}\right) \left( X_{\nu )}^{\mathrm{N}}+\frac{i}{2}\partial _{\nu )}^{\mathrm{%
N}}\right) \phi (X) \\
&=&\left( X_{\mu }\cdot X_{\nu }+\frac{i}{2}D_{\mu \nu }-\frac{1}{4}\partial
_{\mu }\cdot \partial _{\nu }\right) \phi (X)\,\,\,\,\,\equiv \,\,\mathcal{D}%
_{\mu \nu }\phi (X) \\
\mathbf{L}^{MN}\star \phi (X) &=&\frac{1}{2}\varepsilon ^{\mu \nu }\left(
X_{\mu }^{\mathrm{[M}}+\frac{i}{2}\partial _{\mu }^{\mathrm{[M}}\right)
\left( X_{\nu }^{\mathrm{N]}}+\frac{i}{2}\partial _{\nu }^{\mathrm{N]}%
}\right) \phi (X) \\
&=&\left( X_{1}^{\mathrm{[M}}X_{2}^{\mathrm{N]}}+\frac{i}{2}D^{\mathrm{MN}}-%
\frac{1}{4}\partial _{1}^{\mathrm{[M}}\partial _{2}^{\mathrm{N]}}\right)
\phi (X)\,\,\equiv \mathcal{D}^{\mathrm{MN}}\phi (X).
\end{eqnarray}
The $\cdot $ refers to contraction of indices with respect to the SO$^{\ast
}\left( D\right) $ metric $\eta _{\mathrm{MN}}.$ Likewise, from
right-multiplications of $\mathbf{Q}_{\mu \nu }$ and $\mathbf{L}^{\mathrm{MN}%
}$ to the function $\phi (X)$, one obtains another set of second-order
differentiations, $\mathcal{\overline{\mathcal{D}}}_{\mu \nu },\mathcal{%
\overline{\mathcal{D}}}^{\mathrm{MN}}$:
\begin{eqnarray}
\phi (X)\,\star \,\mathbf{Q}_{\mu \nu } &=&\,\,\left( X_{\mu }\cdot X_{\nu }-%
\frac{i}{2}D_{\mu \nu }-\frac{1}{4}\partial _{\mu }\cdot \partial _{\nu
}\right) \phi (X)\,\,\equiv \mathcal{\overline{\mathcal{D}}}_{\mu \nu }\phi
(X) \\
\phi (X)\star \mathbf{L}^{\mathrm{MN}} &=&\left( X_{1}^{[M}X_{2}^{N]}-\frac{i%
}{2}D^{\mathrm{MN}}-\frac{1}{4}\partial _{1}^{[M}\partial _{2}^{N]}\right)
\phi (X)\equiv \mathcal{\overline{\mathcal{D}}}^{\mathrm{MN}}\phi (X).
\end{eqnarray}

These various first and second-order left- and right- differential operators
violate the Leibniz rule, however, they have interesting properties: from
the commutation relations of $\mathbf{Q}_{\mu \nu }$ and $\mathbf{L}^{%
\mathrm{MN}}$, Eqs.(\ref{spso2}-\ref{spso1}), it follows immediately that
each of the sets of differential operators we have defined $\left( D_{\mu
\nu },D^{\mathrm{MN}},\partial _{\mu }^{\mathrm{M}}\right) $ or $\left(
\mathcal{D}_{\mu \nu },\mathcal{D}^{\mathrm{MN}},\mathcal{D}_{\mu }^{\mathrm{%
M}}\right) $ or $\left( \mathcal{\overline{\mathcal{D}}}_{\mu \nu },\mathcal{%
\overline{\mathcal{D}}}^{MN},\mathcal{\overline{\mathcal{D}}}_{\mu }^{%
\mathrm{M}}\right) $ provide inequivalent representations for the generators
of the Sp$\left( 2\right) \times $SO$^{\ast }\left( D\right) $ symmetry
group, as they obey the sp$(2)\oplus $ so${}^{\ast }(D)$ Lie algebra.
\begin{eqnarray}
\Big[\,\mathcal{D}_{\mu \nu }\,,\,\mathcal{D}_{\kappa \lambda }\,\Big] &=&i{%
F_{\mu \nu ,\kappa \lambda }}^{\alpha \beta }\,\,\mathcal{D}_{\alpha \beta },
\label{sp2structure} \\
\left[ \mathcal{D}^{\mathrm{MN}},\mathcal{D}^{\mathrm{KL}}\right] &=&i{F^{%
\mathrm{MN,KL}}}_{\mathrm{PQ}}\mathcal{D}^{\mathrm{PQ}}, \\
\left[ \mathcal{D}^{\mathrm{MN}},\,\mathcal{D}_{\mu \nu }\right] &=&0,
\label{spso4}
\end{eqnarray}
and rotate the first order derivatives in the appropriate fundamental
representation
\begin{eqnarray}
\left[ \,\mathcal{D}_{\mu \nu }\,,\mathcal{D}_{\lambda }^{\mathrm{K}}\right]
&=&i\varepsilon _{\nu \lambda }\mathcal{D}_{\mu }^{\mathrm{K}}+i\varepsilon
_{\mu \lambda }\mathcal{D}_{\mu }^{\mathrm{K}}, \\
\left[ \mathcal{D}^{\mathrm{MN}},\mathcal{D}_{\lambda }^{\mathrm{K}}\right]
&=&i\eta ^{\mathrm{NK}}\mathcal{D}_{\lambda }^{\mathrm{M}}-i\eta ^{\mathrm{MK%
}}\mathcal{D}_{\lambda }^{\mathrm{N}}
\end{eqnarray}
Similar commutation relations are obeyed by the other sets of differential
operators $\left( D_{\mu \nu },D^{\mathrm{MN}},\partial _{\mu }^{\mathrm{M}%
}\right) $ or $\left( \mathcal{\overline{\mathcal{D}}}_{\mu \nu },\mathcal{%
\overline{\mathcal{D}}}^{MN},\mathcal{\overline{\mathcal{D}}}_{\mu }^{%
\mathrm{M}}\right) .$

There also exists another class of second-order differential operators of
the form $X_{\mu }^{\mathrm{M}}\star \phi (X)\star X_{\nu }^{\mathrm{N}}$'s.
One can show, however, that their algebra does not close among themselves
and hence is not relevant for the representation of Sp(2)$\times $SO$%
{}^{\ast }(D)$ symmetry group.

Summarizing the result of this section, we have constructed various first-
and second-order differential operators. The Leibniz rule is obeyed by $%
\partial _{\mu }^{\mathrm{M}}$,$D_{\mu \nu }$,$D^{\mathrm{MN}}$ and violated
by $\mathcal{D}_{\mu }^{\mathrm{M}}$,$\mathcal{\overline{\mathcal{D}}}_{\mu
}^{\mathrm{M}}$, $\mathcal{D}_{\mu \nu }$,$\overline{\mathcal{\mathcal{D}}}%
_{\mu \nu }$, $\mathcal{D}^{\mathrm{MN}}$, $\mathcal{\overline{\mathcal{D}}}%
^{\mathrm{MN}}$. Nevertheless, in the following discussion, all of them will
play a role.


\subsection{Projective Relations}

Extending further the products the field with higher powers of $X_{\mu }^{%
\mathrm{M}}$'s, consider $\star $-multiplication between fields. Given a set
of fields that are well-defined on phase-space, satisfying a suitable
fall-off condition at infinity, the $\star $-multiplication between them
ought to correspond to another field well-defined on the same phase-space,
viz.
\begin{equation}
\star \qquad :\quad \mathcal{M}_{\theta }\otimes \mathcal{M}_{\theta }\qquad
\longrightarrow \qquad \mathcal{M}_{\theta }.  \label{orthogonality}
\end{equation}
We will define a complete basis of fields that close under the $\star $%
-product, and will prove Eq.(\ref{orthogonality}) via explicit calculation.
Recall that, in the context of non-relativistic quantum mechanics, the
Wigner function defined on the particle's phase-space \cite{wigner} is the
Weyl-Moyal counterpart of the \textsl{diagonal} density-matrix operators. We
will begin with generalizing this correspondence to a complete set of
covariant fields defined on relativistic phase-space by including \textsl{%
off-diagonal} density-matrix operators.

Consider a complete set of covariant fields, $\varphi _{m}\left(
X_{1}\right) \equiv <X_{1}|\varphi _{m}>,$ $m=1,2,3,\cdots $, defined on the
particle's configuration-space, and construct all possible density matrices $%
\hat{\rho}_{mn}:=|\varphi _{m}\rangle \langle \varphi _{n}|$ out of them.
Then, noncommutative scalar fields $\phi _{mn}(X_{1},X_{2})$ can be defined
by applying the Weyl-Moyal map to the density matrix:
\begin{eqnarray}
\phi _{mn}\left( X_{1},X_{2}\right) &:&=\int d^{D}Y\,\,\varphi
_{m}(X_{1})\star \,\exp \left( -iX_{2}\cdot Y\right) \star \varphi
_{n}^{\ast }(X_{1}),  \label{wignerr} \\
&=&\int d^{D}Y\,\,\varphi _{m} \left(X_{1}-\frac{Y}{2} \right) \exp \left(
-iX_{2}\cdot Y\right) \varphi _{n}^{\ast } \left(X_{1}+\frac{Y}{2} \right)
\end{eqnarray}
The phase-space field $\phi _{mn}(X)$ is nothing but the Wigner
transformation \cite{wigner} of the configuration space fields $\varphi
_{m}(X_{1}),\varphi _{n}(X_{1})$, now extended to a relativistically
covariant and off-diagonal form. Assuming completeness, one can construct a
\textsl{coherent superposition} to represent any noncommutative field in the
form
\begin{equation}
\phi (X_{1},X_{2}):=\sum C_{mn}\,\phi _{mn}(X_{1},X_{2}).  \label{superpose}
\end{equation}
where $C_{mn}$ are a set of constant coefficients. Therefore it is useful to
learn about the properties of the $\phi _{mn}.$

We claim that noncommutative fields of the form Eq.(\ref{wignerr}) form a
set that close under the $\star $-multiplication, as in Eq.(\ref
{orthogonality}). Explicitly, consider two noncommutative fields, $\phi
_{k\ell }(X),\phi _{mn}(X)$, of the form Eq.(\ref{wignerr}) and take the $%
\star $-product between them. One calculates that
\begin{eqnarray}
\left( \phi _{k\ell }\star \phi _{mn}\right) (X) &=&\int d^{D}Yd^{D}%
\widetilde{Y}\,\left( \varphi _{k}(X_{1})\star e^{-iY\cdot X_{2}}\star
\varphi _{\ell }^{\ast }(X_{1})\right) \star \left( \varphi _{m}(X_{1})\star
e^{-i\tilde{Y}\cdot X_{2}}\star \varphi _{n}^{\ast }(X_{1})\right)  \nonumber
\\
&=&\int d^{D}Yd^{D}\widetilde{Y}\,\varphi _{k}(X_{1})\star \left[
e^{-iY\cdot X_{2}}\star \left( \varphi _{\ell }^{\ast }(X_{1})\varphi
_{m}(X_{1})\right) \star e^{-i\tilde{Y}\cdot X_{2}}\right] \star \varphi
_{n}^{\ast }(X_{1})  \nonumber \\
&=&2^{-D}\int d^{D}Y_{+}\,\varphi _{k}(X_{1})\star \left[ e^{-iX_{2}\cdot
Y_{+}}\int d^{D}Y_{-} \varphi _{\ell }^{\ast }\left( X_{1}- {\frac{ Y_{-} }{2%
}} \right)\varphi _{m}\left(X_{1}- {\frac{ Y_{-} }{2}} \right) \right] \star
\varphi _{n}^{\ast }(X_{1}).  \nonumber
\end{eqnarray}
where $Y_{\pm }^{\mathrm{M}}=\left( Y^{\mathrm{M}}\pm \tilde{Y}^{\mathrm{M}%
}\right) .$ In going from the second to the third line we used the fact that
under the $\star $-product, phase-space `plane-waves', $e^{-ia\cdot X}\equiv
\exp (-ia_{i}^{\mathrm{M}}X_{j}^{\mathrm{N}}\varepsilon ^{ij}\eta_{\mathrm{MN%
}})$, generate translation on the quantum phase-space: for any function $%
F(X_{i}^{\mathrm{M}})$ on phase-space,
\begin{equation}
e^{-ib\cdot X}\star F(X_{i}^{\mathrm{M}})\star e^{-ia\cdot X}=e^{-iX\cdot
(a+b)}\,F\left( X_{i}^{\mathrm{M}}-{\frac{1}{2}}a_{i}^{\mathrm{M}}+{\frac{1}{%
2}}b_{i}^{\mathrm{M}}\right) .
\end{equation}
Since the integrals over $Y_{\pm }^{\mathrm{M}}$ are factorized, one finally
obtains
\begin{equation}
(\phi _{k\ell }\star \phi _{mn})(X)=\mathcal{N}_{\ell m}\,\int
d^{D}Y_{+}\,\varphi _{k}(X_{1})\star e^{-iX_{2}\cdot Y_{+}}\star \varphi
_{n}^{\ast }(X_{1}),
\end{equation}
where $\mathcal{N}_{\ell m}$ is a constant
\begin{eqnarray}
\mathcal{N}_{\ell m} &=&2^{-D}\int d^{D}Y_{-}\,\varphi _{\ell }^{\ast
}(X_{1}-Y_{-}/2)\,\varphi _{m}(X_{1}-Y_{-}/2) \\
&=&\int d^{D}X_{1}\,\varphi _{\ell }^{\ast }(X_{1})\varphi _{m}(X_{1})
\end{eqnarray}
which denotes inner-product between two configuration-space fields, or
simply $\mathcal{N}_{\ell m}=\langle \varphi _{l}|\varphi _{m}\rangle .$
Thus the closure of the algebra satisfied by the $\phi _{mn}$ under $\star $%
-products is the same as the one satisfied by density matrices $\hat{\rho}%
_{mn}:=|\varphi _{m}\rangle \langle \varphi _{n}|.$

In case the configuration-space fields $\varphi _{m}$'s form an orthonormal
basis -- take, for example, configuration space plane-waves, $e^{iX_{1}\cdot
K}$ --, viz. $\mathcal{N}_{\ell m}=\delta _{\ell ,m}$. One then obtains
covariant version of the orthogonality relation:
\begin{equation}
\left( \phi _{k\ell }\star \phi _{mn}\right) (X)=\delta _{\ell ,m}\phi
_{kn}(X)
\end{equation}
as the fundamental nonlinear relations among the noncommutative fields. A
subset closed under the orthogonality relation consists of \textsl{diagonal}
noncommutative fields $\phi _{mm}(X)$, which have the property of projection
operators $|\varphi _{m}\rangle \langle \varphi _{m}|,$ satisfying:
\begin{equation}
\left( \phi _{mm}\star \cdots \star \phi _{mm}\right) (X)=\phi _{mm}(X).
\end{equation}
In fact, such a projection operator $\phi _{mm}$ is a relativistic
generalization of the Wigner distribution function:
\begin{eqnarray}
\phi _{mm} &=&\int d^{D}Y\,\varphi _{m}(X_{1})\star \exp \left( -iX_{2}\cdot
Y\right) \star \varphi _{m}^{\ast }(X_{1}) \\
&=&\int d^{D}Y\,\varphi _{m}\left( X_{1}-Y/2\right) \exp (-iX_{2}\cdot
Y)\,\varphi _{m}\left( X_{1}+Y/2\right) .
\end{eqnarray}
In the solution of our NCFT equations we will use the general superposition (%
\ref{superpose}) to relate 2T-physics in noncommutative field theory to
2T-physics in configuration space as discussed in the following subsection.

Incidentally, in recent works on noncommutative solitons \cite{gopa}, both
diagonal and off-diagonal Wigner distribution functions have been utilized.
Interpreting the $D\times 2$-dimensional phase-space as $D\times 2$%
-dimensional noncommutative space, diagonal Wigner functions are interpreted
as spherically symmetric solitons, while off-diagonal ones are interpreted
as asymmetric solitons. Indeed, the two are related each other by U($\infty $%
) transformations.


\subsection{Map Between Phase-Space and Configuration Space}

Consider the Fourier transform in the $X_{2}$ variable of the general field
in NCFT
\begin{equation}
\phi \left( X_{1},X_{2}\right) :=\int d^{D}Y\,\,\exp \left( -iX_{2}\cdot
Y\right) \,F \left(X_{1}-\frac{Y}{2},X_{1}+\frac{Y}{2} \right)
\label{fourier}
\end{equation}
where $F(X_{L}^{\mathrm{M}},X_{R}^{\mathrm{M}}):=f\left( X_{1},Y\right) $ is
a by-local field in \textit{configuration space}. If one computes the $\star
$-products $X_{1}^{\mathrm{M}}\star \phi \left( X_{1},X_{2}\right) $ and $%
X_{2}^{\mathrm{M}}\star \phi \left( X_{1},X_{2}\right) $ acting from the
left as in (\ref{leftX}), then their effect is reproduced by acting only on
the left variable in $F(X_{L}^{\mathrm{M}},X_{R}^{\mathrm{M}})$ like
position and derivative in configuration space respectively. A similar
result is obtained by acting from the right
\begin{eqnarray}
X_{1}^{\mathrm{M}}\star \phi \left( X_{1},X_{2}\right) \quad &\rightarrow
&\quad X_{L}^{\mathrm{M}}\,\ F(X_{L}^{\mathrm{M}},X_{R}^{\mathrm{M}}),
\label{x1L} \\
X_{2}^{\mathrm{M}}\star \phi \left( X_{1},X_{2}\right) \quad &\rightarrow
&\quad -i\frac{\partial }{\partial X_{L}^{\mathrm{M}}}\,\ F(X_{L}^{\mathrm{M}%
}, X_{R}^{\mathrm{M}}), \\
\phi \left( X_{1},X_{2}\right) \star X_{1}^{\mathrm{M}}\quad &\rightarrow
&\quad X_{R}^{\mathrm{M}}\,\ F(X_{L}^{M},X_{R}^{\mathrm{M}}), \\
\phi \left( X_{1},X_{2}\right) \star X_{2}^{\mathrm{M}}\quad &\rightarrow
&\quad i\frac{\partial }{\partial X_{R}^{\mathrm{M}}}\,\ F(X_{L}^{\mathrm{M}%
}, X_{R}^{\mathrm{M}}).  \label{x2R}
\end{eqnarray}
The left-hand side of these equations is equal to the Fourier transform of
the right hand side as in Eq.(\ref{fourier}). Similarly, we may consider the
basis $\phi _{mn}\left( X_{1},X_{2}\right) $ of the previous section. From
their definition Eq.(\ref{wignerr}) we see that $F(X_{1}-\frac{Y}{2},X_{1}+%
\frac{Y}{2})$ is replaced by $\varphi _{m}(X_{1}-\frac{Y}{2})\,\varphi
_{n}^{\ast }\left( X_{1}+\frac{Y}{2}\right) ,$ and assuming the completeness
of the superposition (\ref{superpose}), we have
\begin{equation}
F(X_{1}-\frac{Y}{2},X_{1}+\frac{Y}{2})=\sum C_{mn}\,\varphi _{m} \left(X_{1}-%
\frac{Y}{2} \right)\,\varphi _{n}^{\ast }\left( X_{1}+\frac{Y}{2}\right) .
\label{Fexpand}
\end{equation}
Using Eq.(\ref{leftX}), we readily verify that the $\star $-multiplication
of $X_{1}^{\mathrm{M}},X_{2}^{\mathrm{M}}$ on the phase-space field $%
\phi_{mn}\left( X_{1},X_{2}\right) $ is equivalent to applying $X_{1}^{%
\mathrm{M}}$ and $-i\partial /\partial X_{1}^{\mathrm{M}}$, respectively, on
configuration-space fields $\varphi _{m}\left( X_{1}^{\mathrm{M}}\right) $.
Explicitly,
\begin{eqnarray}
X_{1}^{\mathrm{M}}\star \phi _{mn}\left( X\right) &=&\int d^{D}Y\,X_{1}^{%
\mathrm{M}}\star \left( \varphi _{m}(X_{1})\star e^{-iX_{2}\cdot Y}\star
\varphi _{n}^{\ast }(X_{1})\right)  \label{x1} \\
&=&\int d^{D}Y\,\Big(X_{1}\varphi _{m}(X_{1})\Big)\star e^{-iX_{2}\cdot
Y}\star \varphi _{n}^{\ast }(X_{1})
\end{eqnarray}
and
\begin{eqnarray}
X_{2}^{\mathrm{M}}\star \phi \left( X\right) &=&\int d^{D}Y\,X_{2}^{\mathrm{M%
}}\star \left( \varphi _{m}(X_{1})\star e^{-iX_{2}\cdot Y}\star \varphi
_{n}^{\ast }(X_{1})\right)  \label{x2} \\
&=&\int d^{D}Y\,\Big(-i\frac{\partial }{\partial X_{1}^{\mathrm{M}}}\varphi
_{m}(X_{1})\Big)\star e^{-iX_{2}\cdot Y}\star \varphi _{n}^{\ast }(X_{1}).
\end{eqnarray}
Therefore, the action $X_{1}^{\mathrm{M}}$ or $X_{2}^{\mathrm{M}}$ on NC
fields, from the left or the right, is equivalent to the usual rules for
position and momenta acting on a complete set of wavefunctions in
configuration space, as illustrated by the expressions in Eqs.(\ref{x1},\ref
{x2}) or in Eqs.(\ref{x1L}-\ref{x2R}).

Using these results, one can show similarly that for the free $\mathbf{Q}%
_{\mu \nu }$ or $\mathbf{L}^{\mathrm{MN}}$ we have
\begin{eqnarray}
\mathbf{Q}_{\mu \nu }\star \phi _{mn}\left( X\right) &=&\int d^{D}Y\,\Big(%
\mathbf{q}_{\mu \nu }\,\varphi _{m}(X_{1})\Big)\star \,e^{-iX_{2}\cdot
Y}\star \varphi _{n}^{\ast }(X_{1}) \\
\mathbf{L}^{\mathrm{MN}}\star \phi _{mn}\left( X\right) &=&\int
d^{D}Y\,\left( \mathbf{l}^{\mathrm{MN}}\,\varphi _{m}(X_{1})\right) \star
e^{-iX_{2}\cdot Y}\star \varphi _{n}^{\ast }(X_{1}),
\end{eqnarray}
where $\mathbf{q}_{\mu \nu }\,\varphi _{m}\left( X_{1}\right) $ and $\mathbf{%
l}^{\mathrm{MN}}\,\varphi _{m}\left( X_{1}\right) $ are given in terms of
ordinary products or derivatives as in Eqs.(\ref{first2}) and (\ref{rot})
respectively. Thus, acting on the basic fields $\varphi _{m}(X_{1})$ in
configuration-space, $\mathbf{q}_{\mu \nu },\,\mathbf{l}^{\mathrm{MN}}$ are
the operators obeying sp$\left( 2\right) \oplus $ so$^{\ast }\left( D\right)
$ Lie algebra:
\begin{eqnarray}
\left[ \mathbf{q}_{\mu \nu },\mathbf{q}_{\alpha \beta }\right] &=&i{F_{\mu
\nu ,\lambda \sigma }}^{\alpha \beta }\,\,\mathbf{q}_{\alpha \beta },
\nonumber \\
\left[ \mathbf{l}^{\mathrm{MN}},\mathbf{l}^{\mathrm{PQ}}\right] &=&i{F^{%
\mathrm{MN,PQ}}}_{\mathrm{RS}}\,\mathbf{l}^{\mathrm{RS}},
\end{eqnarray}
an immediate consequence of Eqs.(\ref{spso2})--(\ref{spso1}). We have seen
in the previous section that these operators have played a prominent role in
the first-quantized approach to 2T-physics.

The above analysis allows us to rewrite the free field equations of
2T-physics in $X_{1}$-space given in Eq.(\ref{first1}) as free field
equations in NCFT in noncommutative phase space
\begin{equation}
\mathbf{q}_{\mu \nu }\,\varphi _{m}\left( X_{1}\right) :=0\quad
\Leftrightarrow \quad \mathbf{Q}_{\mu \nu }\star \phi \left(
X_{1},X_{2}\right) =0=\phi \left( X_{1},X_{2}\right) \star \mathbf{Q}_{\mu
\nu }  \label{qQ}
\end{equation}
A complete set of solutions to the free NCFT equations is provided by a
complete set of solutions to the configuration space free field equations.
These were already solved in \cite{field2T}. Thus the 2T- to 1T-holographic
properties of 2T-physics in NC phase space are closely related to those of
the configuration space by the above map. The complete set of states $%
\varphi _{m}\left( X_{1}\right) $ form a specific unitary representation of
SO$\left( D-2,2\right) $ with quadratic Casimir $D\left( 4-D\right) /4,$
namely the singleton/doubleton, as explained in the paragraphs following Eq.(%
\ref{spso}). Hence the noncommutative field $\phi \left( X_{1},X_{2}\right) $
that satisfies the NC free field equation should be regarded as the direct
product of two singletons/doubletons.

For the more general 2T-physics theory in the presence of background fields,
the first quantized field equation (\ref{fieldeqs}) can also be rewritten
simply in the noncommutative field theory approach as
\begin{equation}
\left( \widehat{\mathbf{Q}}_{\mu \nu }\star \phi \right) \left(
X_{1},X_{2}\right) =0=\left( \phi \star \widehat{\mathbf{Q}}_{\mu \nu
}\right) \left( X_{1},X_{2}\right)  \label{matter}
\end{equation}
where $\widehat{\mathbf{Q}}_{\mu \nu }\left( X_{1},X_{2}\right) $ contains
all background fields, including scalar, vector (gauge field), tensor
(gravitational field) and higher spin fields as analyzed in \cite{highspin}.
The $\widehat{\mathbf{Q}}_{\mu \nu }\left( X_{1},X_{2}\right) $ are required
to obey the Sp$\left( 2\right) $ Lie algebra using star products since at
the classical level they had to obey the same algebra using Poisson brackets
\begin{equation}
\left[ \widehat{\mathbf{Q}}_{\mu \nu },\widehat{\mathbf{Q}}_{\lambda \sigma }%
\right] _{\star }=i\left( \varepsilon _{\nu \lambda }\widehat{\mathbf{Q}}%
_{\mu \sigma }+\varepsilon _{\mu \lambda }\widehat{\mathbf{Q}}_{\nu \sigma
}+\varepsilon _{\nu \sigma }\widehat{\mathbf{Q}}_{\mu \lambda }+\varepsilon
_{\mu \sigma }\widehat{\mathbf{Q}}_{\nu \lambda }\right) .  \label{Sp2}
\end{equation}
Having established the desired field equations in NCFT, including background
fields (before adding further non-linear interactions among the NC fields $%
\phi ,\widehat{\mathbf{Q}}_{\mu \nu }$), we will next proceed to developing
the methodology for deriving them from first principles directly in the NCFT
setting. This requires a study of both global and local Sp(2)$\times $ SO$%
{}^{\ast }(D)$ covariance in NCFT.


\section{Field Theory with Global Sp(2)$\times$ SO${}^*(D)$ Symmetry}

Having identified the symmetry group on relativistic phase-space, we next
construct noncommutative field theory, in which the Sp(2) symmetry is
global. Since this is a new subject which may have more general
applications, we will first develop some general methodology before
returning to the 2T-physics problem.

We begin with specifying Sp$\left( 2\right) \times $ SO${}^{\ast }\left(
D\right) $ representations to the noncommutative fields. The generators that
act on the relativistic phase-space are $\mathbf{Q}_{\mu \nu }$ and $\mathbf{%
L}^{\mathrm{MN}}$, Eqs.(\ref{QandL},\ref{Q&L}). Noncommutative fields of
different defining representations are specified, depending on whether the
generators act on fields from the left, the right, or as a commutator.
Additionally, the fields can carry $\mu ,\nu ,\cdots ;\mathrm{M,N},\cdots $
or spinor indices, thus describing states of higher-spin in Sp$\left(
2\right) $ or SO$^{\ast }\left( D\right) $.

\subsection{`Adjoint' Representations}

By a noncommutative scalar field $\phi \left( X\right) $ in `adjoint'
representation of Sp(2)$\times $SO${}^{\ast }(D)$, we refer to the
transformation rules:
\begin{equation}
\delta _{\mathrm{sp}}\phi (X)=-\frac{i}{2}\omega ^{\alpha \beta }\left[
\mathbf{Q}_{\alpha \beta },\phi \right] _{\star }(X)=\frac{1}{2}\omega
^{\alpha \beta }D_{\alpha \beta }\phi (X),  \label{dsp}
\end{equation}
under infinitesimal Sp(2) rotation, where $\omega ^{\alpha \beta }$ denote
three infinitesimal rotation parameters, and
\begin{equation}
\delta _{\mathrm{so}}\phi (X)=-\frac{i}{2}\omega _{\mathrm{MN}}\left[
\mathbf{L}^{\mathrm{MN}},\phi \right] _{\star }(X)=\frac{1}{2}\omega _{%
\mathrm{MN}}D^{\mathrm{MN}}\phi (X),  \label{dso}
\end{equation}
under infinitesimal SO${}^{\ast }(D)$ rotation, where $\omega _{\mathrm{MN}}$
denote $D\left( D-1\right) $ infinitesimal parameters of SO$^{\ast }\left(
D\right) $ rotations. Note that the latter transformation involves the total
``angular momentum'' operator $X_{1}^{\mathrm{[M}}\partial ^{\mathrm{1N]}%
}+X_{2}^{\mathrm{[M}}\partial ^{\mathrm{2N]}}$, rotating both $X_{1}^{%
\mathrm{M}}$ and $X_{2}^{\mathrm{M}}$ coordinates of the relativistic
phase-space.

From Eq.(\ref{dsp}), one finds Sp(2) transformation rules for various
differential operators acting on the scalar field. Explicitly,
\begin{eqnarray}
\delta _{\mathrm{sp}}\left( \partial _{\mu }^{\mathrm{M}}\phi \right) (X)
&=&\partial _{\mu }^{\mathrm{M}}\left( \frac{1}{2}\omega ^{\alpha \beta
}D_{\alpha \beta }\phi \right) (X) \\
&=&\frac{1}{2}\omega ^{\alpha \beta }D_{\alpha \beta }\left( D_{\mu }^{%
\mathrm{M}}\phi \right) (X)+{\omega _{\mu }}^{\beta }\left( \partial _{\beta
}^{\mathrm{M}}\phi \right) (X).
\end{eqnarray}
The second term arises because $\partial _{\mu }^{\mathrm{M}}\phi $
transforms as Sp(2) doublet as opposed to $\phi (X)$ itself being Sp(2)
singlet. Similarly, $D_{\mu \nu }\phi ,$ $\mathcal{D}_{\mu \nu }\phi ,$ $%
\overline{\mathcal{D}}_{\mu \nu }\phi $ transforms as Sp$\left( 2\right) $
triplets, while $D^{\mathrm{MN}}\phi ,$ $\mathcal{D}^{\mathrm{MN}}\phi ,$ $%
\overline{\mathcal{D}}^{\mathrm{MN}}\phi $ transforms as Sp$\left( 2\right) $
singlets. Hence,
\begin{eqnarray}
\delta _{\mathrm{sp}}\left( D_{\mu \nu }\phi \right) (X) &=&\frac{1}{2}%
\omega ^{\alpha \beta }D_{\alpha \beta }\left( D_{\mu \nu }\phi \right) (X)+{%
\omega _{\mu }}^{\beta }\left( D_{\beta \nu }\phi \right) (X)+{\omega _{\nu }%
}^{\beta }\left( D_{\beta \mu }\phi \right) (X), \\
\delta _{\mathrm{sp}}\left( D^{\mathrm{MN}}\phi \right) (X) &=&\frac{1}{2}%
\omega ^{\alpha \beta }D_{\alpha \beta }\left( D^{\mathrm{MN}}\phi \right)
(X),
\end{eqnarray}
and similarly for $\left( \mathcal{D}_{\mu \nu }\phi ,\,\mathcal{D}^{\mathrm{%
MN}}\phi \right) $ or $\left( \overline{\mathcal{D}}_{\mu \nu }\phi ,\,%
\overline{\mathcal{D}}^{\mathrm{MN}}\phi \right) $. For an infinitesimal SO$%
^{\ast }\left( D\right) $ rotation, transformation rules are obtained
analogously: $D^{\mathrm{MN}}\phi $, $\mathcal{D}^{\mathrm{MN}}\phi $, $%
\overline{\mathcal{D}}^{\mathrm{MN}}\phi $ transform as SO${}^{\ast }(D)$
adjoints, while $D_{\mu \nu }\phi $, $\mathcal{D}_{\mu \nu }\phi $, $%
\overline{\mathcal{D}}_{\mu \nu }\phi $ transforms as SO${}^{\ast }(D)$
singlets.

Having identified the adjoint Sp(2) and SO${}^{\ast }(D)$ transformation
rules, we now proceed to the construction of an action functional possessing
manifest global Sp$\left( 2\right) \times $SO$^{\ast }\left( D\right) $
invariance.

We begin with the potential term. Consider an arbitrary $\star $-product
polynomial of $\phi $'s:
\begin{equation}
V_{\star }(\phi )=m^{2}\phi \star \phi +{\frac{\lambda _{3}}{3}}\phi \star
\phi \star \phi +{\frac{\lambda _{4}}{4}}\phi \star \phi \star \phi \star
\phi +\cdots .
\end{equation}
Eqs.(\ref{dsp}, \ref{dso}) and cyclicity of the $\star $-multiplication then
imply that its integral is invariant under the Sp(2)$\times $SO${}^{\ast
}(D) $ transformations. Explicitly,
\begin{eqnarray}
\delta _{\mathrm{sp}}\left( \,\int d^{2D}X\,V_{\star }(\phi )\right) \,
&=&\int d^{2D}X\,V_{\star }^{\prime }(\phi )\star \delta _{\mathrm{sp}}\phi
=-{\frac{i}{2}}\omega ^{\alpha \beta }\int d^{2D}X\,\Big[\mathbf{Q}_{\alpha
\beta },V_{\star }(\phi )\Big]_{\star }=0, \\
\delta _{\mathrm{so}}\left( \int d^{2D}X\,V_{\star }(\phi )\right) &=&\int
d^{2D}X\,V_{\star }^{\prime }(\phi )\star \delta _{\mathrm{so}}\phi =-{\frac{%
i}{2}}\omega _{\mathrm{LM}}\int d^{2D}X\,\left[ \mathbf{L}^{\mathrm{MN}%
},V_{\star }(\phi )\right] _{\star }=0,
\end{eqnarray}
where cyclicity property of the $\star $-multiplication is used.

Consider next the kinetic term. Possible terms quadratic in differential
operators are given by:
\begin{eqnarray}
\frac{1}{2}\left( \partial _{\mathrm{M}}^{\mu }\phi \right) \star \left(
\partial _{\mu }^{\mathrm{M}}\phi \right) \quad &,&\quad \frac{1}{2}\left(
\overline{\mathcal{D}}_{\mathrm{M}}^{\mu }\phi \right) \star \left( \mathcal{%
D}_{\mu }^{\mathrm{M}}\phi \right) \\
\quad \frac{1}{4}\left( D^{\mu \nu }\phi \right) \star \left( D_{\mu \nu
}\phi \right) \quad &,&\quad \frac{1}{4}\left( D^{\mathrm{MN}}\phi \right)
\star \left( D_{\mathrm{MN}}\phi \right) \\
\frac{1}{4}\left( \overline{\mathcal{D}}^{\mu \nu }\phi \right) \star \left(
\mathcal{D}_{\mu \nu }\phi \right) &=&{\frac{1}{4}}\phi \star \left(
\mathcal{D}^{\mu \nu }\mathcal{D}_{\mu \nu }\phi \right) , \\
\frac{1}{4}\left( \overline{\mathcal{D}}^{\mathrm{MN}}\phi \right) \star
\left( \mathcal{D}_{\mathrm{MN}}\phi \right) &=&{\frac{1}{4}}\phi \star
\left( \mathcal{D}^{\mathrm{MN}}\mathcal{D}_{\mathrm{MN}}\phi \right) .
\end{eqnarray}
All indices are raised or lowered by the Sp$\left( 2\right) $ or SO$^{\ast
}\left( D\right) $ metrics, $\varepsilon _{\mu \nu }$ or $\eta^{\mathrm{MN}}$%
. Because of that, the integrals of the two terms in the first line vanish
identically. The rest, which will be denoted collectively as $\mathcal{L}_{%
\mathrm{KE}}$, all behave as scalars under Sp$\left( 2\right) \times $SO$%
^{\ast }\left( D\right) $ transformations. Hence, like the potential term,
their integrals are invariant once the cyclicity of the $\star $%
-multiplication is taken into account:
\begin{eqnarray}
\delta _{\mathrm{sp}}\int d^{2D}X\,\mathcal{L}_{\mathrm{KE}} &=&-{\frac{i}{2}%
}\,\omega ^{\alpha \beta }\int d^{2D}X\,\,\Big[\mathbf{Q}_{\alpha \beta },%
\mathcal{L}_{\mathrm{KE}}\Big]_{\star }=0, \\
\delta _{\mathrm{so}}\int d^{2D}X\,\mathcal{L}_{\mathrm{KE}} &=&-{\frac{i}{2}%
}\omega _{\mathrm{MN}}\int d^{2D}X\left[ \mathbf{L}^{\mathrm{MN}},\mathcal{L}%
_{\mathrm{KE}}\right] _{\star }=0.
\end{eqnarray}
Furthermore, because of the relation Eq.(\ref{L2Q2}), the last two terms are
related each other:
\begin{equation}
\frac{1}{2}\phi \star \mathcal{D}^{\mathrm{MN}}\mathcal{D}_{\mathrm{MN}}\phi
\,=\,\frac{1}{2}\phi \star \mathcal{D}^{\mu \nu }\mathcal{D}_{\mu \nu }\phi +%
\frac{1}{4}D(4-D)\phi \star \phi .  \label{relation}
\end{equation}
Overall, the most general Sp(2)$\times $SO${}^{\ast }(D)$ invariant action
functional of the `adjoint' scalar field is expressible as:
\begin{equation}
I[\phi ]=\int d^{2D}X\,\left[ \frac{a}{4}\left( \overline{\mathcal{D}}^{\mu
\nu }\phi \right) \star \left( \mathcal{D}_{\mu \nu }\phi \right) +\frac{b}{4%
}\left( D^{\mu \nu }\phi \right) \star \left( D_{\mu \nu }\phi \right) +%
\frac{c}{4}\left( D^{\mathrm{MN}}\phi \right) \star \left( D_{\mathrm{MN}%
}\phi \right) -V_{\star }\left( \phi \right) \right] ,
\end{equation}
where $a,b,c$ denote arbitrary constants.

Inclusion of fermions is straightforward. Denote SO$^{\ast }\left( D\right) $
spinors as $\psi _{\alpha }\left( X\right) .$ The Sp$\left( 2\right) $
invariant differential operators are extendible to the spinors. By
contracting them with SO${}^{\ast }(D)$ Dirac matrices, one obtains possible
kinetic terms as:
\begin{equation}
iD_{\mathrm{MN}}\left( \Gamma ^{\mathrm{MN}}\psi \right) _{\alpha },\qquad i%
\mathcal{D}_{\mathrm{MN}}\left( \Gamma ^{\mathrm{MN}}\psi \right) _{\alpha
},\qquad i\overline{\mathcal{D}}_{\mathrm{MN}}\left( \Gamma ^{\mathrm{MN}%
}\psi \right) _{\alpha }.
\end{equation}
As an example, consider a fermion $\psi $ interacting with a scalar field $%
\phi $, all transforming in `adjoint' representation under Sp(2). The Sp(2)$%
\times $SO${}^{\ast }(D)$ invariant action is then given by:
\begin{equation}
I[\overline{\psi },\psi ,\phi ]=\int d^{2D}X\,\left[ i\overline{\psi }\star
\gamma ^{\mathrm{MN}}\,\left( a^{\prime }\mathcal{D}_{\mathrm{MN}}+b^{\prime
}\overline{\mathcal{D}}_{\mathrm{MN}}+c^{\prime }D_{\mathrm{MN}}\right) \psi
+g\overline{\psi }\star \phi \star \psi +\cdots \right] ,
\end{equation}
where $a^{\prime },b^{\prime },c^{\prime }$ are arbitrary constants and $g$
denotes the Yukawa coupling parameter.


\subsection{`Fundamental' Representations}

In the previous section, we have shown that left or right multiplication of $%
\mathbf{Q}_{\alpha \beta }$'s and $\mathbf{L}^{\mathrm{MN}}$'s yield, in
addition to commutator multiplication, another representations of the sp(2) $%
\oplus $ so${}^{\ast }(D)$ Lie algebra. Based on this, we define left- or
right-`fundamental' representation of a noncommutative scalar field $\Phi
(X) $ by the following transformation rules:
\begin{eqnarray}
\delta _{\mathrm{sp}}^{L}\Phi (X) &=&+\frac{i}{2}\omega _{L}^{\alpha \beta }%
\mathbf{Q}_{\alpha \beta }\star \Phi (X):=+\frac{i}{2}\omega _{L}^{\alpha
\beta }\,\left( \mathcal{D}_{\alpha \beta }\Phi \right) (X),
\label{fundleft} \\
\delta _{\mathrm{sp}}^{R}\Phi (X) &=&-\frac{i}{2}\Phi (X)\star \mathbf{Q}%
_{\alpha \beta }\,\omega _{R}^{\alpha \beta }:=-\frac{i}{2}\left( \overline{%
\mathcal{D}}_{\alpha \beta }\Phi \right) (X)\,\omega _{R}^{\alpha \beta },
\label{fund1transf}
\end{eqnarray}
where $\omega _{L}^{\alpha \beta },\omega _{R}^{\alpha \beta }$ denotes
infinitesimal Sp$\left( 2\right) _{L}$ and Sp$\left( 2\right) _{R}$
transformation parameters. Note that the field $\Phi $ ought to be
complex-valued, in contrast to the `adjoint' representation scalar $\phi $,
which could be real- or complex-valued. Then, the hermitian conjugate field
transforms as
\begin{eqnarray}
\delta _{\mathrm{sp}}^{L}\Phi ^{\dagger }(X) &=&-\frac{i}{2}\left( \Phi
^{\dagger }(X)\star \mathbf{Q}_{\alpha \beta }\right) \,\omega _{L}^{\alpha
\beta }=-\frac{i}{2}\left( \overline{\mathcal{D}}_{\alpha \beta }\Phi
^{\dagger }\right) (X)\,\omega _{L}^{\alpha \beta }, \\
\delta _{\mathrm{sp}}^{R}\Phi ^{\dagger }(X) &=&+\frac{i}{2}\omega
_{R}^{\alpha \beta }\,\left( \mathbf{Q}_{\alpha \beta }\star \Phi ^{\dagger
}(X)\right) =+\frac{i}{2}\omega _{R}^{\alpha \beta }\,\left( \mathcal{D}%
_{\alpha \beta }\Phi ^{\dagger }\right) (X).  \label{fund2transf}
\end{eqnarray}
Likewise, for SO${}^{\ast }(D)$ transformation, left- or right-`fundamental'
representations can be defined analogously.

From Eqs.(\ref{fund1transf}, \ref{fund2transf}), it also follows that $\Phi
\star \Phi ^{\dagger }$ and $\Phi ^{\dagger }\star \Phi $ transform as:
\begin{eqnarray}
\delta _{\mathrm{sp}}^{L}\left( \Phi \star \Phi ^{\dagger }\right) (X) &=&%
\frac{1}{2}\omega _{L}^{\alpha \beta }D_{\alpha \beta }\left( \Phi \star
\Phi ^{\dagger }\right) (X)\quad \mathrm{and}\quad \delta _{\mathrm{sp}%
}^{R}\left( \Phi \star \Phi ^{\dagger }\right) (X)=0, \\
\delta _{\mathrm{sp}}^{L}\left( \Phi ^{\dagger }\star \Phi \right) (X)
&=&0\quad \mathrm{and}\quad \delta _{\mathrm{sp}}^{R}\left( \Phi ^{\dagger
}\star \Phi \right) (X)=\frac{1}{2}\omega _{R}^{\alpha \beta }D_{\alpha
\beta }\left( \Phi ^{\dagger }\star \Phi \right) (X).
\end{eqnarray}
Note that the infinitesimal transformations of $\Phi \star \Phi ^{\dagger }$
are all given entirely in terms of the $D_{\alpha \beta }:=-i\left[ \mathbf{Q%
}_{\alpha \beta },\,\circ \right] $, the derivation satisfying the Leibniz
rule, although the transformation of $\Phi $ involves $\overline{\mathcal{D}}%
_{\alpha \beta },$ the differential operator which does not satisfy the
Leibniz rule. It then follows that any function of $\Phi \star \Phi
^{\dagger }$, $V_{\star }\left( \Phi \star \Phi ^{\dagger }\right) $, is
invariant under Sp$(2)_{R}$ and transforms as an `adjoint' representation
under Sp$(2)_{L}$. Thus,
\begin{equation}
\delta _{\mathrm{sp}}^{R}V_{\star }\left( \Phi \star \Phi ^{\dagger }\right)
=0\quad \mathrm{and}\quad \delta _{\mathrm{sp}}^{L}V_{\star }\left( \Phi
\star \Phi ^{\dagger }\right) =\frac{1}{2}\omega _{L}^{\alpha \beta
}D_{\alpha \beta }V_{\star }\left( \Phi \star \Phi ^{\dagger }\right)
\end{equation}
and vice versa for any function of $\Phi ^{\dagger }\star \Phi $, $V_{\star
}(\Phi ^{\dagger }\star \Phi )$. Therefore, taking $V_{\star }(\Phi
^{\dagger }\star \Phi )$ or $V_{\star }(\Phi \star \Phi ^{\dagger })$ as the
potential term, its integral is invariant manifestly under both Sp$%
(2)_{L}\times $ SO${}^{\ast }(D)_{L}$ and Sp$(2)_{R}\times $ SO${}^{\ast
}(D)_{R}$ transformations.

Next, to construct a kinetic term in the action integral, consider various
differential operators acting on the fields $\Phi ,\Phi ^{\dagger }$. Begin
with $D_{\alpha \beta }\Phi $ and $D^{\mathrm{MN}}\Phi $. Under Sp$(2)_{L}$
transformation
\begin{equation}
\delta _{\mathrm{sp}}^{L}\left( D_{\alpha \beta }\Phi \right) (X)=\omega
_{L}\cdot \mathcal{D}\left( D_{\alpha \beta }\Phi \right) (X)-\left( {\omega
_{L}}\cdot \mathcal{D}\right) _{(\alpha \beta )}\Phi (X)
\end{equation}
generates the $\mathcal{D}$ differential operator in the second term, hence $%
D_{\mu \nu }\Phi $ is not covariant under Sp$\left( 2\right) _{L}.$
Analogous results apply for Sp$(2)_{R}$, SO${}^{\ast }(D)_{L}$, and SO$%
{}^{\ast }(D)_{R}$ transformations. Hence $D_{\mu \nu }\Phi $ and $D^{%
\mathrm{MN}}\Phi $ do not define covariant differential operators, contrary
to the situation for the fields in `adjoint' representation. It turns out
that Sp$(2)_{L}$ and Sp$(2)_{R}$ covariant differential operators are given
by $\mathcal{D}_{\mu \nu }\Phi $ and $\overline{\mathcal{D}}_{\mu \nu }\Phi $%
, respectively. Explicitly,
\begin{eqnarray}
\delta _{\mathrm{sp}}^{L}\left( \mathcal{D}_{\alpha \beta }\Phi \right) (X)
&=&+{\frac{i}{2}}\omega _{L}\cdot \mathcal{D}\left( \mathcal{D}_{\alpha
\beta }\Phi \right) (X)-\left( \omega _{L}\cdot \mathcal{D}\right) _{(\alpha
\beta )}\Phi (X) \\
\delta _{\mathrm{sp}}^{R}\left( \overline{\mathcal{D}}_{\alpha \beta }\Phi
\right) (X) &=&-{\frac{i}{2}}\omega _{R}\cdot \mathcal{D}\left( \overline{%
\mathcal{D}}_{\alpha \beta }\Phi \right) (X)-\left( \omega _{R}\cdot
\overline{\mathcal{D}}\right) _{(\alpha \beta )}\Phi (X)
\end{eqnarray}
and analogous expressions for SO${}^{\ast }(D)_{L,R}$ transformations for $%
\mathcal{D}^{\mathrm{MN}}\Phi $ and $\overline{\mathcal{D}}^{\mathrm{MN}%
}\Phi $ differential operators. Thus, $\mathcal{D}_{\alpha \beta }\Phi $, $%
\overline{\mathcal{D}}_{\alpha \beta }\Phi $, $\mathcal{D}^{\mathrm{MN}}\Phi
$, and $\overline{\mathcal{D}}^{\mathrm{MN}}\Phi $ transform as adjoint
representations under Sp$(2)_{L}$, Sp$(2)_{R}$, SO${}^{\ast }(D)_{L}$, and SO%
${}^{\ast }(D)_{R}$, respectively, and as singlets otherwise. Similarly, $%
\mathcal{D}_{\mu }^{M}\Phi $ transforms in the fundamental representation of
Sp$\left( 2\right) _{L}\times $SO$^{\ast }\left( D\right) _{L}$ and in the
singlet of Sp$\left( 2\right) _{R}\times $SO$^{\ast }\left( D\right) _{R},$
while $\overline{\mathcal{D}}_{\mu }^{\mathrm{M}}\Phi $ transforms in the
singlet of Sp$\left( 2\right) _{L}\times $SO${}^{\ast }\left( D\right) _{L}$
and in the fundamental representation of Sp$\left( 2\right) _{R}\times $SO $%
{}^{\ast }\left( D\right) _{R}.$ The differential operators acting on the
hermitian conjugate field $\Phi ^{\dagger }$ exhibit similar transformation
rules, related to those of $\Phi $ by interchanging the left- and the
right-symmetry groups.

Putting the above results together, for the `fundamental' scalar field $\Phi
$, the most general action integral with manifest Sp$(2)_{L}\times $ Sp$%
(2)_{R}\times $ SO${}^{\ast }(D)_{L}\times $ SO${}^{\ast }(D)_{R}$
invariance is given by
\begin{equation}
I[\Phi ,\Phi ^{\dagger }]=\int d^{2D}X\left[ a\,\overline{\mathcal{D}}^{\mu
\nu }\Phi ^{\dagger }\star \mathcal{D}_{\mu \nu }\Phi +b\,\mathcal{D}_{\mu
\nu }\Phi ^{\dagger }\star \overline{\mathcal{D}}^{\mu \nu }\Phi -V_{\star
}\left( \Phi \star \Phi ^{\dagger }\right) \right] .
\end{equation}
One could have also added terms of the form $\overline{\mathcal{D}}^{\mathrm{%
MN}}\Phi ^{\dagger }\star \mathcal{D}_{\mathrm{MN}}\Phi $ and $\mathcal{D}^{%
\mathrm{MN}}\Phi ^{\dagger }\star \overline{\mathcal{D}}_{\mathrm{MN}}\Phi $%
. As pointed out in Eq.(\ref{relation}), they are re-expressible in terms of
those already included. In the action, $a,b$ are arbitrary coefficients.

The field equation of motion is given by:
\begin{equation}
{\frac{1}{2}}\left( a\mathcal{D}^{\mu \nu }\mathcal{D}_{\mu \nu }+b\overline{%
\mathcal{D}}^{\mu \nu }\overline{\mathcal{D}}_{\mu \nu }\right) \Phi
=V_{\star }^{\prime }\left( \Phi \star \Phi ^{\dagger }\right) \star \Phi .
\end{equation}
Note that the left-hand side is expressed entirely in terms of the Sp$\left(
2\right) _{L}$ and Sp$\left( 2\right) _{R}$ Casimir operators, viz. ${\frac{1%
}{2}}\mathcal{D}^{\mu \nu }\mathcal{D}_{\mu \nu }={\frac{1}{2}}\mathbf{Q}%
^{\mu \nu }\star \mathbf{Q}_{\mu \nu }={\frac{1}{2}}\overline{\mathcal{D}}%
^{\mu \nu }\overline{\mathcal{D}}_{\mu \nu }$ acting on $\Phi $ either from
the left or from the right.

Extension to fermion or higher-rank tensor field is straightforward. The
fermion $\Psi _{\alpha }$ can be taken as the spinor representation of
either SO$^{\ast }\left( D\right) _{L}$ or SO$^{\ast }\left( D\right) _{R}$
and as the `fundamental' representation of either Sp$(2)_{L}$ or Sp$(2)_{R}$%
. Taking, as an example, that $\Psi _{\alpha }$ is in the
left-representations for both, the action integral is expressible as:
\begin{equation}
I[\overline{\Psi },\Psi ]=\int d^{2D}X\,\left[ \overline{\Psi }\star \left(
\Gamma ^{\mathrm{MN}}\cdot i\mathcal{D}_{\mathrm{MN}}\Psi \right) +\cdots %
\right] .
\end{equation}
The ellipses denote interaction part, whose form is constrained severely by
the requirement of both the Sp$(2)_{L}\times $ SO${}^{\ast }(D)_{L}$ and the
Sp$(2)_{R}\times $ SO${}^{\ast }(D)_{R}$ symmetry groups.


\subsection{Spacetime Signature and Automorphism Group}

We have constructed noncommutative field theories on the relativistic
phase-space, in which the phase-space Sp$(2)\times $ SO${}^{\ast }(D)$ is
manifest. Because Sp(2) is part of the manifest symmetry group, $X_{1},X_{2}$
appear explicitly in the kinetic terms, and break translation symmetry.
These field theories are SO${}^{\ast }(D)$ Lorentz invariant but not
Poincar\'{e} invariant in $D$ dimensions. It implies, in particular, energy
and momentum cannot be used as quantum numbers labelling states in Hilbert
space. The good quantum numbers are associated with the representations of Sp%
$(2)\times $ SO${}^{\ast }(D).$

Is this an indication that something is wrong with the theory? Not at all.
The lack of translation invariance is a common feature of 2T-physics in all
its formulations, and surprisingly it is correct from the lower dimensional
1T physical point of view. When we identify the 1T-dynamics in the lower $%
(D-2)$ dimensions, the system does have Poincar\'{e} symmetry from the $D-2$
dimensional point of view. An example in one of the holographic pictures is
that SO${}^{\ast }(D)=$SO$\left( D-2,2\right) $ is the conformal group in $%
D-2$ dimensions, and it does contain the Poincar\'{e} group, including
translations. This example shows that embedding the symmetries of the
physical space in SO$^{\ast }\left( D\right) $ is possible, and that the
embedding space may have some unusual signature.

In more general cases beyond 2T-physics, the spacetime signature, which has
been left unspecified so far, ought to be determined by consistency and
physical properties of the theory. There are several ways of doing so. One
is by treating the spacetime coordinates $X_{1}^{\mathrm{M}}$'s as embedding
space coordinates of a true physical spacetime as in the 2T-physics example.
For instance, one may formulate Euclidean quantum field theories on a $(D-1)$%
-dimensional hypersphere in terms of those on $D$-dimensional Euclidean
space \cite{adler}. Likewise, quantum field theories on $(D-2)$-dimensional
de Sitter space can be recasted in terms of those on a $D$-dimensional
Lorentzian spacetime with \textsl{one} timelike dimension, and quantum field
theories on $(D-2)$-dimensional anti-de Sitter space in terms of those on a $%
D$-dimensional Lorentzian spacetime with \textsl{two} timelike dimensions.
In all cases, the physical spacetime is defined as a hypersurface defined by
an appropriate quadratic equations for coordinates of the $D$-dimensional
embedding space. Moreover, the symmetry group of the physical spacetime is SO%
${}^{\ast }(D)$ and acts linearly on coordinates of the embedding space. Any
of these embeddings will require some local symmetry to thin out degrees of
freedom, eliminate ghosts, and reduce the theory to the lower dimensional
theory.

The above discussion suggests that the noncommutative field theory with
global Sp(2)$\times $SO${}^{\ast }(D)$ automorphism group may be viewed as a
sort of theory defined on an embedding phase-space of the physical
phase-space. In particular, the signature of the higher dimensional
spacetime will be determined depending on the way the physical phase-space
is embedded into the higher dimensional space.

\section{Field Theory with Local Sp$(2)$ Symmetry}

In this section, we will discuss noncommutative Sp(2) gauge theory on
relativistic noncommutative phase-space. Of particular interest would be the
construction of a theory, whose field equations coincide with Eqs.(\ref
{matter},\ref{Sp2}) for 2T-physics.

\subsection{Action and equations of motion}

Consider promoting the global Sp(2)$_{L}$ transformation Eq.( \ref{fundleft}%
) of the complex scalar field $\Phi $, to a local transformation
parametrized by $\omega _{L}^{\alpha \beta }\left( X_{1},X_{2}\right) $:
\begin{equation}
\delta ^{L}\Phi (X)=\frac{i}{2}\omega ^{\alpha \beta }(X)\star \left(
\mathcal{D}_{\alpha \beta }\Phi \right) (X)=\frac{i}{2}\left( \omega
^{\alpha \beta }(X)\star \mathbf{Q}_{\alpha \beta }\right) \star \Phi
(X)\equiv i\omega _{L}(X)\star \Phi (X).  \label{dLf}
\end{equation}
Ordering of the factors in $\frac{1}{2}\omega _{L}^{\alpha \beta }(X)\star
\mathbf{Q}_{\alpha \beta }:=\omega _{L}\left( X\right) $ could be more
general. With any ordering, the resulting $i\omega _{L}\left(
X_{1},X_{2}\right) $ can be regarded as the general noncommutative
infinitesimal local phase transformation acting on the left of $\Phi $. So,
we will in fact interpret local Sp$\left( 2\right) _{L}$ applied on a scalar
to mean the general gauge transformation for any $\omega _{L}\left( X\right)
$ applied from the left as in the last expression in Eq.( \ref{dLf}).
Proceeding as usual, we introduce a gauge potential $\mathbf{A}_{\mu \nu
}\left( X_{1},X_{2}\right) $ and promote the global Sp$\left( 2\right) _{L}$
differential operator $\mathcal{D}_{\mu \nu }$ to a local covariant
differential operator $\widehat{\mathcal{D}}_{\mu \nu }$%
\begin{equation}
\widehat{\mathcal{D}}_{\mu \nu }\Phi (X):=\mathcal{D}_{\mu \nu }\Phi (X)+%
\mathbf{A}_{\mu \nu }\star \Phi (X)=\Big(\mathbf{Q}_{\mu \nu }+\mathbf{A}%
_{\mu \nu }(X)\Big)\star \Phi (X).  \label{covder}
\end{equation}
The noncommutative local transformations are defined by Eq.(\ref{dLf}) along
with
\begin{equation}
\delta ^{L}\mathbf{A}_{\mu \nu }(X)=\widehat{D}_{\mu \nu }\,\omega
_{L}=D_{\mu \nu }\,\omega _{L}(X)-i\left[ \mathbf{A}_{\mu \nu }(X),\omega
_{L}(X)\right] _{\star }=-i\left[ \left( \mathbf{Q}_{\mu \nu }+\mathbf{A}%
_{\mu \nu }\right) ,\omega _{L}\right] _{\star },  \label{deltaA}
\end{equation}
where $D_{\mu \nu }$ is the derivation of Eq.(\ref{Dij}) that satisfies the
Leibniz rule. This ensures the covariance of the differential operator $\hat{%
\mathcal{D}}_{\mu \nu }\Phi $:
\begin{equation}
\delta ^{L}\left( \widehat{\mathcal{D}}_{\mu \nu }\Phi \right) =i\omega
_{L}\star \widehat{\mathcal{D}}_{\mu \nu }\Phi .
\end{equation}
Denote the covariantized $\mathbf{Q}_{\mu \nu }$ as $\widehat{\mathbf{Q}}%
_{\mu \nu }\left( X_{1},X_{2}\right) $
\begin{equation}
\widehat{\mathbf{Q}}_{\mu \nu }:=\mathbf{Q}_{\mu \nu }+\mathbf{A}_{\mu \nu }=%
\frac{1}{2}X_{(\mu }^{\mathrm{M}}\star X_{\nu )}^{\mathrm{N}}\eta _{\mathrm{%
MN}}+\mathbf{A}_{\mu \nu }\left( X\right) .
\end{equation}
Note that $\widehat{\mathbf{Q}}_{\mu \nu }\left( X_{1},X_{2}\right) $ is the
counterpart of the classical $\hat{Q}_{\mu \nu }\left( X,P\right) $ that
appeared in the worldline formalism in Eq.(\ref{worldline}). The
infinitesimal local gauge transformation of Eq.(\ref{deltaA}) is
re-expressed as
\begin{equation}
\delta ^{L}\widehat{\mathbf{Q}}_{\mu \nu }=-i\left[ \widehat{\mathbf{Q}}%
_{\mu \nu },\omega _{L}\right] _{\star }.  \label{deltaQ}
\end{equation}
This is the counterpart of the canonical transformations in the ``space of
all theories'' discussed in the worldline approach \cite{highspin}.

The covariant field strength $\mathbf{G}_{\mu \nu ,\lambda \sigma }^{L}(X)$
is obtained from the $\star $-commutator of the covariant derivatives:
\begin{eqnarray}
\left[ \widehat{\mathcal{D}}_{\mu \nu },\widehat{\mathcal{D}}_{\lambda
\sigma }\right] _{\star }\star \Phi (X) &=&\Big[\left( \mathcal{D}_{\mu \nu
}+\mathbf{A}_{\mu \nu }\right) ,\left( \mathcal{D}_{\lambda \sigma }+\mathbf{%
A}_{\lambda \sigma }\right) \Big]_{\star }\,\star \Phi (X) \\
&=&i\,{F_{\mu \nu ,\lambda \sigma }}^{\alpha \beta }\,\left( \mathcal{D}%
_{\alpha \beta }\Phi \right) +i\left( D_{\mu \nu }\mathbf{A}_{\lambda \sigma
}-D_{\lambda \sigma }\mathbf{A}_{\mu \nu }-i\left[ \mathbf{A}_{\mu \nu },%
\mathbf{A}_{\lambda \sigma }\right] _{\star }\right) \star \Phi  \\
&=&i\,{F_{\mu \nu ,\lambda \sigma }}^{\alpha \beta }\,\widehat{\mathcal{D}}%
_{\alpha \beta }\Phi +i\,\mathbf{G}_{\mu \nu ,\lambda \sigma }\star \Phi ,
\label{ff}
\end{eqnarray}
where ${F_{\mu \nu ,\lambda \sigma }}^{\alpha \beta }$ refers to
the Sp$ \left( 2\right) $ structure constants,
Eq.(\ref{structures}), and the covariant field strength is given
by
\begin{equation}
\mathbf{G}_{\mu \nu ,\lambda \sigma }(X)=D_{\mu \nu }\mathbf{A}_{\lambda
\sigma }-D_{\lambda \sigma }\mathbf{A}_{\mu \nu }-i\Big[\mathbf{A}_{\mu \nu
},\mathbf{A}_{\lambda \sigma }\Big]_{\star }-i\,{F_{\mu \nu ,\lambda \sigma }%
}^{\alpha \beta }\mathbf{A}_{\alpha \beta }.  \label{fieldstrength}
\end{equation}
Note again that $D_{\mu \nu }$ is the derivation of Eq.(\ref{Dij}) that
satisfies the Leibniz rule. The last term in the field strength originates
from the covariantization of the non-Abelian differential operators
involved. In terms of the covariant generators $\widehat{\mathbf{Q}}_{\mu
\nu }(X)$, the field strength becomes
\begin{equation}
i\,\mathbf{G}_{\mu \nu ,\lambda \sigma }=\left[ \widehat{\mathbf{Q}}_{\mu
\nu },\widehat{\mathbf{Q}}_{\lambda \sigma }\right] _{\star }-i\,{F_{\mu \nu
,\lambda \sigma }}^{\alpha \beta }\,\widehat{\mathbf{Q}}_{\alpha \beta }.
\label{Fieldstrength}
\end{equation}
$\mathbf{G}_{\mu \nu ,\lambda \sigma }$ has only three independent
components which may be rewritten in the form of a symmetric 2$\times $2
tensor $\mathbf{G}^{\mu \nu },$ the latter being obtained from contraction
of $\mathbf{G}_{\lambda \sigma ,\rho \sigma }$ with the structure constant
raised indices ${F}^{\mu \nu ,\lambda \sigma ,\rho \sigma }.$ Explicitly,
three independent components of the $\mathbf{G}^{\mu \nu }$ takes the form
\begin{eqnarray}
\mathbf{G}^{11} &=&i\left[ \widehat{\mathbf{Q}}_{12},\widehat{\mathbf{Q}}%
_{22}\right] _{\star }+2\widehat{\mathbf{Q}}_{22},  \label{Gij} \\
\mathbf{G}^{12} &=&\frac{i}{2}\left[ \widehat{\mathbf{Q}}_{22},\widehat{%
\mathbf{Q}}_{11}\right] _{\star }-2\widehat{\mathbf{Q}}_{12}, \\
\mathbf{G}^{22} &=&i\left[ \widehat{\mathbf{Q}}_{11},\widehat{\mathbf{Q}}%
_{12}\right] _{\star }+2\widehat{\mathbf{Q}}_{11}.  \label{G22}
\end{eqnarray}

The vanishing of the field strengths $\mathbf{G}^{\mu \nu }$ or $\mathbf{G}%
_{\mu \nu ,\lambda \sigma }$ is equivalent to $\widehat{\mathbf{Q}}_{\mu \nu
}$ satisfying the first quantized sp$\left( 2\right) $ algebra as in Eq.(\ref
{Sp2}). This algebra had emerged as a condition in the first quantized
worldline theory Eq.(\ref{worldline}), which followed from the identical
algebra in the form of Poisson brackets in the classical theory. Thus, we
now aim at deriving the equations $\mathbf{G}^{\mu \nu }=0$ as equations of
motion (before possible field interactions) from an action principle in the
noncommutative field theory. We can easily obtain this result from the
following noncommutative field theory, whose structure is analogous to the
Chern-Simons gauge theory:
\begin{eqnarray}
S_{Q}&=& \int d^{2D} X \, \left[ \left< \widehat{\mathbf{Q}}, \widehat{%
\mathbf{Q}} \star \widehat{\mathbf{Q}} \right> - \left< \widehat{\mathbf{Q}}%
, \widehat{\mathbf{Q}} \right> \right]  \nonumber \\
&:=& \int d^{2D}X\,\left(
\begin{array}{c}
i \widehat{\mathbf{Q}}_{11}\star \widehat{\mathbf{Q}}_{12}\star \widehat{%
\mathbf{Q}}_{22}-i\widehat{\mathbf{Q}}_{22}\star \widehat{\mathbf{Q}}%
_{12}\star \widehat{\mathbf{Q}}_{11} \\
+\widehat{\mathbf{Q}}_{11}\star \widehat{\mathbf{Q}}_{22}+\widehat{\mathbf{Q}%
}_{22}\star \widehat{\mathbf{Q}}_{11}-2\widehat{\mathbf{Q}}_{12}\star
\widehat{\mathbf{Q}}_{12}
\end{array}
\right) ,  \label{action1}
\end{eqnarray}
whose variation yields $\delta S_{Q}=\int d^{2D}X\,\left( \delta \widehat{%
\mathbf{Q}}_{\mu \nu }\star \mathbf{G}^{\mu \nu }\right) .$

To obtain also the equation Eq.(\ref{matter}) for the matter field $\Phi
\left(X_{1},X_{2}\right)$, consider the covariant derivative Eq.(\ref{covder}%
), $\widehat{\mathcal{D} }_{\mu \nu } \Phi \equiv \widehat{\mathbf{Q} }_{\mu
\nu}\star \Phi $, and add it to the action Eq.(\ref{action1}) after
multiplying it with a Lagrange multiplier field $Z^{\mu \nu }\left(
X_{1},X_{2}\right) $:
\begin{equation}
S_{\Phi ,Q,Z}=-i\int d^{2D}X\left( \overline{Z}^{\mu \nu }\star \widehat{%
\mathbf{Q}}_{\mu \nu }\star \Phi -\overline{\Phi}\star \widehat{\mathbf{Q}}%
_{\mu \nu }\star Z^{\mu \nu }\right).  \label{action2}
\end{equation}
The $\overline{Z}^{\mu \nu }$ field equation yields the free part of the the
desired matter equation
\begin{equation}
\widehat{\mathbf{Q}}_{\mu \nu }\star \Phi =0,
\end{equation}
while $\overline{\Phi}$ field equation yields an equation for $Z^{\mu \nu }$%
of the form
\begin{equation}
\widehat{\mathbf{Q}}_{\mu \nu }\star Z^{\mu \nu }=0.
\end{equation}
The action $S_{\Phi ,Q,Z}$ is invariant under the local Sp$(2)_L$
transformations Eqs.(\ref{dLf},\ref{deltaQ}) provided $Z^{\mu \nu }$ field
transfoms as $\delta ^{L}Z^{\mu \nu }=i\omega _{L}\star Z^{\mu \nu }$, and
under the local Sp$\left( 2\right) _{R}$ defined by
\begin{equation}
\delta ^{R}\Phi =-i\Phi \star \omega _{R}\qquad \mathrm{and} \qquad \delta
^{R}Z^{\mu \nu }=-iZ^{\mu \nu }\star \omega _{R}.
\end{equation}
One may accordingly define a Hermitian field $\phi \left( X_{1},X_{2}\right)
=\Phi \star \overline{\Phi}$ satisfying $\widehat{\mathbf{Q}}_{\mu \nu
}\star \phi =0=\phi \star \widehat{\mathbf{Q}}_{\mu \nu }$, corresponding to
the first-quantized matter wavefunction of the worldline theory, Eq.(\ref
{matter}).

The addition of matter fields would give rise to a back-reaction to the
gauge fields themselves. The field equations derived from the combined
action
\begin{equation}
S_{\mathrm{total}} =S_{Q}+S_{\Phi ,Q,Z}  \label{SQS}
\end{equation}
are
\begin{equation}
\mathbf{G}^{\mu \nu }=\Phi \star \overline{Z}^{\mu \nu }-Z^{\mu \nu } \star
\overline{\Phi},\quad \widehat{\mathbf{{Q} }}_{\mu \nu }\star \Phi =0,\qquad
\widehat{\mathbf{{Q} }}_{\mu \nu }\star Z^{\mu \nu }=0,  \label{fieldeqns}
\end{equation}
plus Hermitian conjugates of the last two equations. From them, one derives
the following field equations involving gauge fields only
\begin{equation}
\widehat{\mathbf{Q}}_{\mu \nu }\star \mathbf{G}^{\mu \nu }=0=\mathbf{G}^{\mu
\nu }\star \widehat{\mathbf{Q}}_{\mu \nu }.  \label{gaugeeqs}
\end{equation}
Evidently, the structure of these equations is consistent with the first
quantization of the worldline theory as given in Eqs.(\ref{Sp2},\ref{matter}%
), in particular, when the matter self interactions are ignored, as then $%
\mathbf{G}^{\mu \nu }=0$ and $\widehat{\mathbf{Q} }_{\mu \nu }\star \Phi =0$%
. One may setup an expansion around this solution and analyze the classical
solution of these equations.

By virtue of the relation to the worldline 2T-physics theory, we are assured
that the spectrum of these 2T-field equations is unitary (ghost-free) and
causal. Indeed, as in the classical theory, the physical spectrum is empty
unless there are two timelike dimensions. Furthermore, the physics described
by them has a direct relation to the 1T-physics in $(D-2)$ dimensions by
virtue of the holographic property of 2T-physics. As demonstrated below (see
also \cite{highspin} ), the $\mathbf{G}^{\mu \nu } \left( X,P\right) =0$
equations describe, when expanded in powers of $P_{\mathrm{M}}$'s,
background gauge fields of various higher--spins in the $(D-2)$-dimensional
spacetime. The matter field equation, the second in Eq.(\ref{fieldeqns}),
implies that these higher-spin fields are coupled also to a scalar field
(the $\varphi \left( X\right) $ in Eq.(\ref{wignerr}) ) in $(D-2)$%
-dimensional spacetime.

Having noted that we have made the desired connection with 2T-physics, one
can generalize the noncommutative Sp(2) gauge theory by including nonlinear
(self)-interactions consistently with gauge and spacetime symmetries. The
inclusion of such interaction, such as Eq.(\ref{gaugeeqs}) and those below,
would generate kinetic terms describing propagation of the gauge fields, but
this has not been studied yet in our setting. This is an interesting issue
for furthre study, as it is related to construction of an \textsl{interacting%
} higher-spin gauge field theory, whose satisfactory solution has remained
elusive despite considerable progress \cite{vasil}. Specifically, consider
adding terms up to two derivatives of $\mathcal{D}_{\mu \nu }$ or $D_{\mu
\nu }.$ Of particular interest would be the Yang-Mills action for the Sp(2)
gauge field $\mathbf{A}_{\mu \nu }(X)$, which can be taken instead of or in
addition to the above Chern-Simons type action:
\begin{eqnarray}
S_{G^{2}} &=& -\frac{1}{4g^{2}}\int d^{2D} X \, \left( \mathbf{G}_{\mu \nu
,\lambda \sigma } \right)^{2}  \nonumber \\
&=& \frac{1}{4g^{2}} \int d^{2D} X \, \left( \left[ \widehat{\mathbf{Q}}%
_{\mu \nu },\widehat{\mathbf{Q}}_{\lambda \sigma }\right] _{\star }-i\,{%
F_{\mu \nu ,\lambda \sigma }}^{\alpha \beta }\widehat{\mathbf{Q}}_{\alpha
\beta }\right) _{\star }^{2}.  \label{ncftaction}
\end{eqnarray}
Similarly, one may add self-interactions of the scalar field $\varphi \left(
X_{1},X_{2}\right) $ (including the scalar field $\Phi $ discussed above):
\begin{eqnarray}
S_{\varphi } &=& \int d^{2D} X \, \left[ -\frac{1}{2}\left( {\widehat{%
\mathcal{D}}}^{\mu \nu }\varphi \right) ^{\dagger }\star \widehat{\mathcal{D}%
}_{\mu \nu }\varphi -V_{\star }\left( \varphi \star \varphi ^{\dagger
}\right) \right]  \nonumber \\
&=& \int d^{2D} X \, \left[ -\frac{1}{2}\varphi ^{\dagger }\star \left(
\widehat{\mathbf{Q}}_{\mu \nu }\right) ^{2}\star \varphi -V_{\star }\left(
\varphi \star \varphi ^{\dagger }\right)\right],  \nonumber
\end{eqnarray}
where $\left( \widehat{\mathbf{Q}}_{\mu \nu }\right) ^{2}$ is the quadratic
Casimir operator of Sp$\left( 2\right).$

\subsection{Classical Solutions}

Let us now analyze physical contents of the equations
\begin{equation}
\mathbf{G}^{\mu \nu }=0\qquad \mathrm{and}\qquad \widehat{\mathbf{Q}}_{\mu
\nu }\star \Phi =0.  \label{physical}
\end{equation}
From Eqs.(\ref{Gij}-\ref{G22}) it is evident that $\mathbf{{G}^{\mu \nu }=0}$
is equivalent to imposing the sp$\left( 2\right) $ algebra on $\widehat{%
\mathbf{Q}}_{\mu \nu }\left( X,P\right) .$ A solution to this situation was
found in \cite{highspin} as follows: using the Sp$(2)_{L}$ gauge
transformations, one can choose gauges such that the $\widehat{\mathbf{Q}}%
_{\mu \nu }\left( X,P\right) $ takes the following form~\footnote{%
We emphasize that after choosing a gauge for $\widehat{\mathbf{Q}}_{11}$,
the remaining gauge symmetry is insufficient to simplify the structure of $%
\widehat{\mathbf{Q}}_{12}\left( X,P\right) $ further. However, if $\widehat{%
\mathbf{Q}}_{12}$ is restricted to obey the sp$\left( 2\right) $ algebra,
the remaining gauge symmetry can be used to set it to the form shown in the
text.}
\begin{eqnarray}
\widehat{\mathbf{Q}}_{11}\left( X,P\right)  &=&X^{\mathrm{M}}X^{\mathrm{N}%
}\eta _{\mathrm{MN}},\quad \widehat{\mathbf{Q}}_{12}\left( X,P\right) =X^{%
\mathrm{M}}\left( P_{\mathrm{M}}+A_{\mathrm{M}}\left( X\right) \right)
\nonumber \\
\widehat{\mathbf{Q}}_{22}\left( X,P\right)  &=&G_{0}\left( X\right) +G_{2}^{%
\mathrm{MN}}\left( P+A\right) _{\mathrm{M}}\left( P+A\right) _{\mathrm{N}}
\nonumber \\
&&+\sum_{s=3}^{\infty }G_{s}^{\mathrm{M_{1}\cdots M_{s}}}\left( X\right)
\left[ \left( P+A\right) _{\mathrm{M_{1}}}\cdots \left( P+A\right) _{\mathrm{%
M_{s}}}\right]   \nonumber
\end{eqnarray}
where $\eta _{\mathrm{MN}}$ is the SO$\left( D-2,2\right) $ metric, $A_{%
\mathrm{M}}\left( X\right) $ is the Maxwell gauge field in $D$ dimensions, $%
G_{0}\left( X\right) $ is the dilaton, $G_{2}^{\mathrm{MN}}\left( X\right)
=\eta ^{\mathrm{MN}}+h_{2}^{\mathrm{MN}}\left( X\right) $ is the spacetime
metric in $D$ dimensions, and $G_{s}^{\mathrm{M_{1}\cdots M_{s}}}\left(
X\right) $ for all $s\geq 3$ are the higher-spin gauge fields. To obey the sp%
$\left( 2\right) $ algebra, these fields ought to be homogeneous polynomials
of degree $(s-2)$ and be orthogonal to $X^{\mathrm{M}}$ (using the flat SO$%
\left( D-2,2\right) $ metric $\eta _{\mathrm{MN}}$) as follows
\begin{equation}
X\cdot \partial G_{s}=\left( s-2\right) G_{s},\qquad X_{\mathrm{M_{1}}%
}G_{s}^{\mathrm{M_{1}\cdots M_{s}}}=X_{\mathrm{M_{1}}}h_{2}^{\mathrm{%
M_{1}M_{2}}}=0,\qquad X^{\mathrm{M}}F_{\mathrm{MN}}=0,  \label{hologr}
\end{equation}
where $F_{\mathrm{MN}}=(\partial _{\mathrm{M}}A_{\mathrm{N}}-\partial _{%
\mathrm{N}}A_{\mathrm{M}})$ is the Maxwell field strength. The Maxwell gauge
symmetry can also be partially fixed by taking $X\cdot A=0$. Then, $X^{%
\mathrm{M}}F_{\mathrm{MN}}=0$ becomes a homogeneity condition $X\cdot
\partial A_{M}=-A_{M}.$ After the gauge-fixing, there still remains local Sp$%
(2)_{L}$ symmetry that does not change the gauge fixed form of the $\widehat{%
\mathbf{Q}}_{11}\left( X,P\right) $ and $\widehat{\mathbf{Q}}_{22}\left(
X,P\right) $ given above (i.e. $\delta ^{L}\widehat{\mathbf{Q}}_{11}=\delta
^{L}\widehat{\mathbf{Q}}_{22}=0$). From Eq.(\ref{deltaQ}), one finds that
the corresponding gauge function $\omega _{L}\left( X,P\right) $ ought to
take the form
\begin{equation}
\omega _{L}\left( X,P\right) =\omega _{0}\left( X\right) +\omega _{1}^{%
\mathrm{M}}\left( X\right) \left( P+A\right) _{\mathrm{M}}+\sum_{s=2}^{%
\infty }\omega _{s}^{\mathrm{M_{1}\cdots M_{s}}}\left( X\right) \left[
\left( P+A\right) _{\mathrm{M_{1}}}\cdots \left( P+A\right) _{\mathrm{M_{s}}}%
\right] ,
\end{equation}
where each coefficient is a homogeneous function of degree $s$ and is
transverse to $X^{\mathrm{M}}$ :
\begin{equation}
X\cdot \partial \omega _{s}=s\,\omega _{s}\qquad \mathrm{and}\qquad X_{%
\mathrm{M_{1}}}\omega _{s}^{\mathrm{M_{1}\cdots M_{s}}}=0\qquad \mathrm{for}%
\qquad s\geq 0.  \label{gaugesymm}
\end{equation}
These residual gauge symmetries are interpreted as follows: $\omega
_{0}\left( X\right) $ is the gauge parameter that transforms the Maxwell
field, $\omega _{1}^{\mathrm{M}}\left( X\right) $ is the parameter for
general coordinate transformations, and the $\omega _{s-1}^{\mathrm{%
M_{1}\cdots M_{s-1}}}$ are gauge parameters for the high spin fields $G_{s}^{%
\mathrm{M_{1}\cdots M_{s}}}$. The gauge transformations mix various gauge
fields one another (see \cite{highspin}), but typically an inhomogeneous
term
\begin{equation}
\delta G_{s}^{\mathrm{M_{1}\cdots M_{s}}}=\partial ^{(M_{1}}\omega _{s-1}^{%
\mathrm{M_{2}\cdots M_{s})}}+\cdots
\end{equation}
remains in the gauge transformations, where the index on $\partial ^{\mathrm{%
M}}$ raised as $\partial ^{\mathrm{M}}=G_{2}^{\mathrm{MN}}\partial _{\mathrm{%
N}}$.

The equations Eqs.(\ref{hologr}), taken together with $X^{2}=0$ (which
equals to the $\widehat{\mathbf{Q}}_{11}=0$ condition), describe fields
whose independent degrees of freedom reside in $(D-2)$ dimensions, both from
the viewpoint of their components and their dependence on spacetime
coordinates. Specifically, Eqs.(\ref{hologr}), together with $X^{2}=0,$
impose the holographic property of 2T-physics. An explicit holographic
projection from $D$ dimensional spacetime $X^{M}$ to $(D-2)$-dimensional
spacetime $x^{\mu }$ is presented in \cite{highspin}. One then sees that the
independent degrees of freedom are given by the fields $g_{0}\left( x\right)
,A_{\mu }\left( x\right) ,$ $g_{\mu \nu }\left( x\right) ,$ $g_{s}^{\mu
_{1}\cdots \mu _{s}}\left( x\right) $ for $s\geq 3,$ which are fields in $%
(D-2)$ dimensions, where the Lorentz components $\mu, \nu, \cdots $
transform according to SO$\left( d-1,1\right) .$ All of these $(D-2)$%
-dimensional fields are consistent with the $(D-2)$-dimensional conformal
symmetry SO$\left( D-2,2\right) $, as this is made evident by the $D$%
-dimensional formalism of 2T-physics.

The remaining gauge symmetries of Eq.(\ref{gaugesymm}) are also
holographically projected to $(D-2)$ dimensions, and their independent
components are $\varepsilon _{0}\left( x\right) ,$ $\varepsilon _{1}^{\mu
}\left( x\right) ,$ and $\varepsilon _{s}^{\mu _{1}\cdots \mu _{s}}\left(
x\right) $ for $s\geq 2.$ It turns out that these remaining gauge symmetries
are strong enough to reduce the fields to pure gauge degrees of freedom,
\textsl{unless} lower- and higher-spin fields do not coexist in the
solution. The exceptional cases therefore lead to two distinct sets of
non-trivial solutions: a \textsl{lower-spin branch} and a \textsl{%
higher-spin branch}. The lower-spin branch consists only of $g_{0}\left(
x\right) ,A_{\mu }\left( x\right) ,$ $g_{\mu \nu }\left( x\right) $, while
all higher-spin fields $\left( s\geq 3\right) $ vanish. In the higher-spin
branch, $g_{0}\left( x\right) ,A_{\mu }\left( x\right) $ vanish, while $%
g_{\mu \nu }\left( x\right) ,$ together with $g_{s}^{\mu _{1}\cdots \mu
_{s}}\left( x\right) $ for $s\geq 3$ form a non-trivial basis for the gauge
transformations, whose explicit forms are calculated in \cite{highspin}.

Intriguingly, the two disconnected branches of solutions appear to bear a
correspondence to massless states of string theories in two extreme limits
(or phases). The lower-spin branch with spins $s\leq 2$ coincides with the
limiting string spectrum in the zero Regge slope limit (infinite tension),
while the higher-spin branch $s\geq 2$ coincides with the limiting string
spectrum of the leading graviton trajectory in the infinite Regge slope
limit (zero tension). \footnote{%
The aforementioned solutions for $\mathbf{G}^{\mu \nu }=0$ describe gauge
fields, but the propagation of these fields is not determined by this
equation. Thus the kinetic term must come from terms such as Eqs.(\ref
{gaugeeqs},\ref{ncftaction}) which have not been included in our
consideration so far.}

We have thus found a set of interesting solutions to Eqs.(\ref{physical})
and have succeeded in their physical interpretations . They have important
implications; the equations $\mathbf{G}^{\mu \nu }=0$ encode \textsl{all}
possible $(D-2)$-dimensional gauge field backgrounds that a spinless
point-particle would interact with. Moreover, the interaction with the
spinless field is governed by the physical state condition, $\widehat{%
\mathbf{Q}}_{\mu \nu }\star \Phi (X)=0,$ -- a condition which solves the
noncommutative field equations when nonlinear interactions are turned off$.$
Via the covariant Wigner transform, Eq.(\ref{wignerr}) and the technology
developed in Eqs.(\ref{x1}-\ref{qQ}), one then obtains the corresponding
field equations for the complete set of fields $\varphi _{m}\left(
X_{1}\right) $ defined on the $D$-dimensional configuration space, but now
in the presence of these background fields. The physical state condition
then reduces them to $(D-2)$-dimensional field equations, again in the
presence of these background fields.

Finally, let us describe how the 2T- to 1T- holography and duality
properties emerge in this formalism. The reduction from $D$-dimensional
spacetime to $(D-2)$-dimensional one has followed from solving the $D$%
-dimensional field equations. The solution can be presented in a variety of
ways of embedding $the (D-2)$ dimensions inside the $D$ dimensions \cite
{field2T}. Different embeddings give rise to different $(D-2)$-dimensional
`holographic' viewpoints of the original $D$-dimensional field equations. In
doing so, which one of the two times becomes \textsl{the} timelike dimension
in the projected $(D-2)$ dimensions? In principle, an infinite number of
choices are available, corresponding to the embedding of a timelike curve in
the extra dimensions. The choice made by the embedding determines the
dynamical evolution of the holographic projection. Each of the $(D-2)$%
-dimensional dynamics may look different, even though any one of them
represents a gauge invariant physical sector of one and the same $D$%
-dimensional theory. This implies that, by a different choice of the Sp$%
\left( 2\right) $ gauge, different $(D-2)$-dimensional theories in different
background fields are obtained and all these theories are transformed one
another by local Sp$\left( 2,R\right) $ gauge transformations. What we have
succeeded in this work is that this property can now be obtained from
\textsl{the first principles} by formulating the 2T-physics in terms of
noncommutative Sp(2) gauge field theories.


\section{Outlook}

In this paper we have constructed noncommutative field theories with global
or local Sp(2) symmetry defined on relativistic phase-space. We believe
these theories deserve further investigation, either as a description of
2T-physics from first principles, or with global Sp(2) symmetry in other
applications.

We mention some of the immediate questions that come up by the results in
this work. First, in noncommutative Sp(2) gauge theories, there is an
important issue concerning gauge-invariant operators. It is known that, in
the context of noncommutative field theories formulated as deformation
quantization over a noncommutative space, part of noncommutative gauge
transformation orbit is identifiable with translation along the
noncommutative space \cite{reyvonunge, dasrey, grosshashimotoitzhaki}. It
implies that gauge-invariant observables are necessarily nonlocal. A
complete set of such observables are identified with open Wilson lines \cite
{reyvonunge, dasrey, grosshashimotoitzhaki}. By a similar argument, in
noncommutative Sp(2) gauge theories formulated in this work, part of
noncommutative gauge transformation orbit ought to be identifiable with
rotation on the relativistic phase-space so that gauge-invariant observables
are nonlocal. We expect that open Wilson lines stretched over the
relativistic phase-space constitute an important class of such observables.
As they are gauge-invariant, from the viewpoint of the two-time physics,
expectation value of the open Wilson lines ought to be universally the same
for all classes of $(D-2)$-dimensional theories related to one another via
the `holography property'. In view of conceptual importance of the latter,
the role of these observables in understanding the `2T- to 1T- holography'
could be extremely rewarding.

Second, a complete classification of noncommutative Sp$\left( 2,R\right) $
gauge theories underlying the 2T-physics is desirable. We have already shown
that a Chern-Simons type action or its variant is a viable route. For this
goal, a BRST approach would offer an economic procedure for construction of
the actions. For example, analogous to Witten's open string field theory,
one can construct a BRST operator
\begin{eqnarray}
\mathcal{Q}_{\mathrm{BRST}} &=&\left\langle \mathbf{c},\widehat{\mathbf{Q}}%
\right\rangle -\left\langle \mathbf{c},\mathbf{c}\star \mathbf{b}%
\right\rangle   \label{qbrst} \\
&\equiv &\mathbf{c}^{\mu \nu }\widehat{\mathbf{Q}}_{\mu \nu }\left(
X,P\right) -\frac{i}{2}{F_{\mu \nu ,\sigma \lambda }}^{\alpha \beta }\,%
\mathbf{c}^{\mu \nu }\mathbf{c}^{\sigma \lambda }\mathbf{b}_{\alpha \beta },
\end{eqnarray}
where $\mathbf{c}^{\mu \nu },\,\mathbf{b}_{\mu \nu }$ are the BRST ghosts
and anti-ghosts, with ghost charge $Q_{\mathrm{gh}}=-1,+1$, respectively.
The ghosts $\mathbf{c}^{\mu \nu }$ and $\mathbf{b}_{\mu \nu }$ represent
three independent fermionic degrees of freedom (one may think of them as 3
creation and 3 annihilation operators acting on fermionic Fock space,
equivalent to six (8$\times $8) matrices with the same anticommutation
properties). There is no need for a definition of star products for the
ghosts (although this is possible via the Weyl correspondence applied to
fermions). Instead of star products they can be treated as fermionic quantum
operators, or Grassmann numbers, keeping track of their orders as usual. We
take an action of the purely cubic Chern-Simons type
\begin{equation}
S_{\mathrm{BRST}}=\int d\mu \lbrack X,\mathbf{b,c}]\,\left( \mathcal{Q}_{%
\mathrm{BRST}}\star \mathcal{Q}_{\mathrm{BRST}}\star \mathcal{Q}_{\mathrm{%
BRST}}\right) ,  \label{brstaction}
\end{equation}
where the star product refers to Moyal product in phase space $\left(
X,P\right) $ we have used in the rest of the paper. The integration measure $%
d\mu \lbrack X,\mathbf{b,c}]=\left( d^{2D}X\right) \,\left( d^{3}\mathbf{c}%
\right) \,\left( d^{3}\mathbf{b}\right) $ $\left( \mathbf{b}_{11}\mathbf{b}%
_{12}\mathbf{b}_{22}\right) $ is invariant under Sp$\left( 2\right) $ and
has ghost number +3, cancelling the ghost number -3 of the Lagrangian
density (instead of fermionic integrals one may also use a vacuum
expectation value in Fock space, or a trace in 8$\times $8 matrix space).
Thus the only term in the Lagrangian that survives the integration is the
term that contains the Sp$\left( 2\right) $ invariant ghost factor $\mathbf{c%
}^{11}\mathbf{c}^{12}\mathbf{c}^{22}..$ Generalizing Eq.(\ref{qbrst}), one
can take the BRST operator $\mathcal{Q}_{\mathrm{BRST}}\left( X,P,\mathbf{c},%
\mathbf{b}\right) $ to be the most general ghost number -1 field, containing
phase space fields as coefficients in all the allowed terms (which have the
form $\mathbf{c},\mathbf{ccb},\mathbf{cccbb}$)$\mathbf{.}$ One may then
define a gauge symmetry on these fields that is given by
\begin{equation}
\delta \mathcal{Q}_{\mathrm{BRST}}=i\left[ \mathcal{Q}_{\mathrm{BRST}%
},\Lambda \right] _{\star },
\end{equation}
where $\Lambda \left( X,P,\mathbf{c},\mathbf{b}\right) $ is a general gauge
function of ghost number zero. Note that, when expanded in powers of ghosts,
the ghost independent term in $\Lambda $ is precisely the local gauge
parameter $\omega _{L}\left( X,P\right) $ discussed earlier. The action, Eq.(%
\ref{brstaction}), is the direct counterpart of the background-independent,
purely cubic action in Witten's open string field theory. Some comparison
points include the fact that $\widehat{\mathbf{Q}}_{\mu \nu }=\mathbf{Q}%
_{\mu \nu }+A_{\mu \nu }$ where $\mathbf{Q}_{\mu \nu }=\frac{1}{2}X_{(\mu
}\cdot X_{\nu )}$ is a particular background, while the general $\mathcal{Q}%
_{\mathrm{BRST}},$ as well as the star product, are background independent.
The equation of motion is $\mathcal{Q}_{\mathrm{BRST}}\star \mathcal{Q}_{%
\mathrm{BRST}}=0$ and, when an appropriate $\Lambda $ gauge is chosen, it
leads to the fundamental equation $\mathbf{G}^{\mu \nu }=0$. This was in the
absence of matter. One may add matter fields $\Psi ,\overline{\Psi }$
containing a linear combination of ghost charges 0,-1,-2, $\Psi =\Psi
_{0}+\Psi _{-1}+\Psi _{-2},$ with an action that takes the form
\begin{equation}
S_{\mathrm{matter}}=\int d\mu \lbrack X,\mathbf{b,c}]\,\left( \overline{\Psi
}\star \mathcal{Q}_{\mathrm{BRST}}\star \Psi \right) .
\end{equation}
The terms that survive integration are those that add up to ghost number -3.
The field equations that follow from the total action $S_{\mathrm{BRST}}+S_{%
\mathrm{matter}}$ are
\begin{equation}
\mathcal{Q}_{\mathrm{BRST}}\star \Psi =0,\qquad \overline{\Psi }\star
\mathcal{Q}_{\mathrm{BRST}}=0,\qquad \mathcal{Q}_{\mathrm{BRST}}\star
\mathcal{Q}_{\mathrm{BRST}}=\left( \Psi \star \overline{\Psi }\right) _{-2},
\end{equation}
where the subscript -2 implies the sum of terms in the product with total
ghost number -2. Thus each matter field $\Psi _{0},\Psi _{-1},\Psi _{-2},$
is annihilated by $\mathcal{Q}_{\mathrm{BRST}}$ separately. These equations
lead to
\begin{equation}
\mathcal{Q}_{\mathrm{BRST}}\star \mathcal{Q}_{\mathrm{BRST}}\star \mathcal{Q}%
_{\mathrm{BRST}}=0,
\end{equation}
which is similar to the nonlinear relation following from the action in Eq.(%
\ref{SQS}). This now looks like an equation of motion for the gauge fields
since it has the form of $\mathcal{Q}_{\mathrm{BRST}}$ (i.e. Klein-Gordon
type operator) applied on $\mathcal{Q}_{\mathrm{BRST}}\star \mathcal{Q}_{%
\mathrm{BRST}}$ (which is like a field strength for the gauge fields).

Given that there are several viable candidate theories, which one would
become eventually `the' proper 2T-physics field theory? A criterion would be
that the kinetic term for the gauge fields ought to be produced correctly by
the proper theory. As computations involving the Moyal star product are
notoriously difficult in the present setting, primarily because they involve
derivatives of all orders, identification of the proper theory would take
considerable effort. We will report progress on this project elsewhere in a
separate paper.

Third, for any given action, further study and a complete classification of
the classical solutions in noncommutative Sp(2) gauge theories are needed.
As uncovered in the present work, classical solutions correspond to variety
of background fields in the holographically projected configuration space.
As such, a complete classification of the classical solutions would lead to
better understanding of many important issues in 2T-physics as well as
1T-physics, in particular, a consistent formulation of interacting
higher-spin field theories \cite{vasil}. We anticipate that classical
solutions with nonzero field strength, $\mathbf{G}^{\mu \nu }(X)\neq 0$, and
nonvanishing scalar self-interactions, $V_{\star }(\Phi \star \Phi ^{\dagger
})\neq 0$, open up new surprises.

Finally, we also expect diverse applications of our formalism and results to
the Euclidean noncommutative field theories arising in string theories and
M-theory \cite{douglas1, douglas2, seibergwitten} and even to other physics
problems than string theories and M-theory.

We will report progress on these issues elsewhere.


\section*{Acknowledgments}

We thank M. Vasiliev and E. Witten for helpful discussions. This work was
initiated during our mutual visits to the CIT-USC Center for Theoretical
Physics and School of Physics at Seoul National University. We thank both
institutions for hospitality. SJR acknowledges warm hospitality of Insitut
Henri Poincar\'{e}, Institut des Hautes \`{E}tudes Scientifique, and The
Institute for Theoretical Physics -- Santa Barbara during the final stage of
this work. I.B. was partially supported by the U.S. DOE grant
DE-FG03-84ER40168, and the CIT-USC Center for Theoretical Physics. S.-J.R.
was partially supported by BK-21 Initiative in Physics (SNU - Project 2),
KRF International Collaboration Grant, KOSEF Interdisciplinary Research
Grant 98-07-02-07-01-5, KOSEF Leading Scientist Program 2000-1-11200-001-1,
and U.S. NSF grant PHY 99-07949.

\end{document}